\documentclass[prd,aps,twocolumn,preprintnumbers, showpacs, nofootinbib,superscriptaddress,notitlepage]{revtex4-1}
\usepackage{amssymb, amsthm}
\usepackage{amsmath}    
\usepackage{mathrsfs}   
\usepackage{graphicx}   
\usepackage{color}      
\usepackage{footnote}   

\usepackage{slashed}    
\usepackage{subfigure}  
\usepackage{hyperref}   

\usepackage{rotating}   
\usepackage{multirow}   
\usepackage[normalem]{ulem}

\newcommand{\non}{\nonumber}

\begin{document}

\preprint{}

\title{Pion generalized parton distribution from lattice QCD}

\author{Jiunn-Wei Chen}
\email{jwc@phys.ntu.edu.tw}
\affiliation{Department of Physics, Center for Theoretical Physics, and Leung Center for Cosmology and Particle Astrophysics, National Taiwan University, Taipei, Taiwan 106}

\author{Huey-Wen Lin}
\email{hwlin@pa.msu.edu}
\affiliation{Department of Physics and Astronomy, Michigan State University, East Lansing, MI 48824}
\affiliation{Department of Computational Mathematics, Michigan State University, East Lansing, MI 48824}

\author{Jian-Hui Zhang}
\email{jianhui.zhang@ur.de}
\affiliation{Institut f\"ur Theoretische Physik, Universit\"at Regensburg, D-93040 Regensburg, Germany}

\begin{abstract}

We present the first lattice calculation of the valence-quark generalized parton distribution (GPD) of the pion using the large-momentum effective theory (LaMET) approach. We focus on the zero-skewness limit, where the GPD has a probability-density interpretation in the longitudinal Bjorken $x$ and the transverse impact-parameter distributions. 
Our calculation is done using clover valence fermions on an ensemble of gauge configurations with $2+1+1$ flavors (degenerate up/down, strange and charm) of highly improved staggered quarks (HISQ) with lattice spacing $a \approx 0.12$~fm, box size $L \approx 3$~fm and pion mass $m_\pi \approx 310$~MeV. The parton distribution function and the form factor are reproduced as special limits of the GPD as expected. Due to the large errors, this exploratory study does not show a clear preference among different model assumptions about the kinematic dependence of the
GPD. To discriminate between these assumptions, future studies using higher-statistics data will be crucial.

\end{abstract}
\maketitle

\section{Introduction}
In the past few decades, extensive studies of parton distribution functions (PDFs) have provided us with detailed knowledge of the longitudinal momentum distribution of quarks and gluons, and, therefore, a one-dimensional picture of hadrons. To map out the multidimensional partonic structure of hadrons, which is an important goal for experiments carried out at the existing facilities in DESY, JLab, BNL, CERN or the planned Electron-Ion Collider, we need to study quantities exhibiting the transverse structure of hadrons. One such quantity that has attracted a lot of interest in the past few years are the generalized parton distributions (GPDs)~\cite{Ji:1996ek,Ji:1996nm} (see also  \cite{Mueller:1998fv}). 

The GPDs unify seemingly different physical quantities, such as the PDFs and hadron form factors, into the same framework. They offer a description of the correlations between the transverse position and
longitudinal momentum of quarks and gluons inside the nucleon, thereby giving access to quark and gluon orbital
angular momentum contributions to the nucleon spin~\cite{Ji:1996ek}. Experimentally, the GPDs can be accessed through hard exclusive processes like deeply virtual Compton scattering or meson production. Useful constraints on the forms of the nucleon GPDs have been obtained from measurements of such processes at DESY~\cite{Airapetian:2001yk,Adloff:2001cn,Chekanov:2003ya} and JLab~\cite{Defurne:2015kxq,Jo:2015ema,Seder:2014cdc,Dudek:2012vr}. However, as the GPDs usually contribute to experimental observables through convolutions and they have more complicated kinematic dependence than the PDFs, extracting the GPDs from these experimental measurements is in general rather difficult. Therefore, inputs from theory are important and play a complementary role in determining the GPDs. Valuable insights are gained through computations using models (see e.g. Ref.~\cite{Belitsky:2005qn} for a review) and lattice QCD. So far, computations using the latter are limited to the first few moments of the GPDs~\cite{Gockeler:2003jfa,Hagler:2003jd,LHPC:2003aa,Gockeler:2005cj} (see Ref.~\cite{Hagler:2009ni} for a review).

In recent years, a new theory framework has been developed that allows for lattice calculations of the $x$-dependence, instead of the moments, of parton quantities~\cite{Ji:2013dva,Ji:2014gla}. This theory is now known as the large-momentum effective theory (LaMET). In this approach, a parton observable such as the PDFs or the GPDs can be accessed from lattice QCD in the following manner: 1) {Construct} an appropriate static-operator matrix element (a quasi-observable) that approaches the parton observable in the large-momentum limit of the external hadron. The quasi-observable constructed this way is usually hadron-momentum--dependent but time-independent, and, therefore, can be readily computed on the lattice. 2) {Calculate} the quasi-observable on the lattice. 3) {Convert} it to the parton observable through a factorization formula accurate up to power corrections that are suppressed by the hadron momentum. The existence of such a factorization is ensured by construction; for a proof, see Refs.~\cite{Ma:2017pxb,Izubuchi:2018srq,Liu:2019urm}.

Since LaMET was proposed, a lot of progress has been achieved with respect to both the theoretical understanding of the formalism~\cite{Xiong:2013bka,Ji:2015jwa,Ji:2015qla,Xiong:2015nua,Ji:2014hxa,Monahan:2017hpu,Ji:2018hvs,Stewart:2017tvs,
Constantinou:2017sej,Green:2017xeu,Izubuchi:2018srq,Xiong:2017jtn,Wang:2017qyg,Wang:2017eel,Xu:2018mpf,
Chen:2016utp,Zhang:2017bzy,Ishikawa:2016znu,Chen:2016fxx,Ji:2017oey,
Ishikawa:2017faj,Chen:2017mzz,Alexandrou:2017huk,Constantinou:2017sej,
Green:2017xeu,Chen:2017mzz,Chen:2017mie,Lin:2017ani,Chen:2017lnm,Li:2016amo,
Monahan:2016bvm,Radyushkin:2016hsy,Rossi:2017muf,Carlson:2017gpk,Ji:2017rah,
Briceno:2018lfj,Hobbs:2017xtq,Jia:2017uul,Xu:2018eii,Jia:2018qee,Spanoudes:2018zya,Rossi:2018zkn,Liu:2018uuj,
Ji:2018waw,Bhattacharya:2018zxi,Radyushkin:2018nbf,Zhang:2018diq,Li:2018tpe,Braun:2018brg} and its application to lattice calculations of nucleon and meson PDFs, as well as meson distribution amplitudes~\cite{Lin:2014zya,Chen:2016utp,Lin:2017ani,Alexandrou:2015rja,Alexandrou:2016jqi,Alexandrou:2017huk,
Chen:2017mzz,Zhang:2017bzy,Chen:2017gck,Alexandrou:2018pbm,Chen:2018xof,Chen:2018fwa,Alexandrou:2018eet,
Lin:2018qky,Fan:2018dxu,Liu:2018hxv,Wang:2019tgg}.
Despite limited volumes and relatively coarse lattice spacings, the state-of-the-art nucleon isovector quark PDFs determined from lattice data at the physical point have shown reasonable agreement~\cite{Chen:2018xof,Lin:2018qky,Alexandrou:2018pbm} with phenomenological results extracted from the experimental data~\cite{Dulat:2015mca,Ball:2017nwa,Harland-Lang:2014zoa,Nocera:2014gqa,Ethier:2017zbq}. Of course, a careful study of theoretical uncertainties and lattice artifacts is still needed to fully establish the reliability of the results.

As for the GPDs, the factorization of the isovector quark quasi-GPDs has been proven to leading-power accuracy using the operator product expansion~\cite{Liu:2019urm}, and the corresponding hard matching function was also computed both in a cutoff scheme~\cite{Ji:2015qla,Xiong:2015nua} and in a regularization-independent momentum-subtraction (RI/MOM) scheme~\cite{Liu:2019urm} (for studies of quasi-GPDs in diquark models see e.g.~\cite{Bhattacharya:2018zxi,Bhattacharya:2019cme}). This allows us to perform exploratory studies on the quark GPDs once we have lattice simulations of the corresponding quasi-GPD matrix elements. 

In this paper, we carry out the first lattice calculation of the valence-quark GPD of the pion using the LaMET approach. As a first step, we focus on the zero-skewness limit, that is, the momentum transfer between the initial and final states is purely transverse. In this limit, the quark GPD is related to the impact-parameter distribution of quarks that has a probability-density interpretation~\cite{Burkardt:2002hr} (see also Ref.~\cite{Ralston:2001xs}).
Our calculation is done using clover valence fermions on an ensemble of gauge configurations with $2+1+1$ flavors (degenerate up/down, strange and charm) of highly improved staggered quarks (HISQ)~\cite{Follana:2006rc} generated by the MILC Collaboration~\cite{Bazavov:2012xda}
with lattice spacing $a \approx 0.12$~fm, box size $L \approx 3$~fm and pion mass $m_\pi \approx 310$~MeV.

\section{From quasi-GPD to GPD in the pion} 

The unpolarized quark GPD in the pion is defined on the lightcone as
\begin{multline}\label{GPD}
H_q^\pi(x,\xi,t,\mu) = \int\frac{d\eta^-}{4\pi} e^{-i x \eta^- P^+} \\
 \times \big\langle \pi(P+\frac{\Delta}{2})\left|{\bar q}(\frac{\eta^-}{2})\gamma^+ \Gamma(\frac{\eta^-}{2}, -\frac{\eta^-}{2})q(-\frac{\eta^-}{2})\right|\pi(P-\frac{\Delta}{2})\big\rangle,
\end{multline}
where $\bar q, q$ denote the quark fields, $\eta^{\pm}=(\eta^0\pm\eta^3)/\sqrt 2$, and $x\in[-1,1]$. {The pion momentum of the initial(final) state is $P \mp \Delta/2$ with $\Delta$ the momentum transfer and $P^\mu=(P^0,0,0,P^z)$. The variables}
\begin{equation}
t=\Delta^2, \hspace{2em} \xi=-\frac{\Delta^+}{2P^+} ,
\end{equation}
{and $\mu$ is the renormalization scale in the $\overline{\text{MS}}$ scheme. The gauge link}
\begin{align}
\Gamma(\eta_2^-,\eta_1^-) &= \exp\bigg(-ig \int_{\eta_1^-}^{\eta_2^-}d\eta^- A^+(\eta^-) \bigg) \,
\end{align}
ensures gauge invariance of the quark bilinear operator.
Eq.~(\ref{GPD}) is an off-forward matrix element where the momenta for the initial and final states are different. In the forward ($\Delta^{\mu} \to 0$) limit, it reduces to the PDF.

An appropriate quark quasi-GPD that can be computed on the lattice is given by
\begin{equation}\label{qGPD}
\tilde{H}_q^\pi(x,\xi,t,P^z,\tilde\mu) =\int\frac{d z P^z}{2\pi} e^{i x P^z z}\tilde h(z, P^z, \xi, t, \tilde\mu)
\end{equation}
with
\begin{multline}\label{qGPDME}
\tilde h(z, P^z, \xi, t, \tilde\mu)=\\
\hspace{-.5em}\frac{1}{2P^0}\big\langle \pi(P+\frac \Delta 2)\left|\bar q(\frac z 2)\gamma^t\Gamma(\frac z 2,-\frac z 2) q(-\frac z 2)\right|\pi(P-\frac \Delta 2)\big\rangle,
\end{multline}
where we have chosen the Dirac matrix as $\gamma^t$, since it has the advantage of avoiding mixing with the scalar quark operator~\cite{Constantinou:2017sej,Chen:2017mie} when a non-chiral lattice fermion is used. This choice will be used throughout this paper. The skewness parameter $\xi$ in Eq.~(\ref{qGPD}) is defined as $\xi=-\Delta^z/(2P^z)$, which differs from the lightcone definition in Eq.~(\ref{GPD}) by power-suppressed contributions of ${\cal O}(m_\pi^2/P_z^2)$. We have ignored this difference and denoted it with the same label as the skewness in the GPD. $\tilde \mu$ denotes the renormalization scale in an appropriate renormalization scheme for the quasi-GPD. In the present paper, we will focus on the $u-d$ combination at zero skewness $\xi=0$, where the former avoids contributions from disconnected diagrams as well as the mixing with gluon GPDs, while the latter simplifies the kinematic dependence of the quark GPD and is also related to the impact-parameter distribution of quarks that has a probability-density interpretation~\cite{Burkardt:2000za,Burkardt:2002hr}. 

The bare pion matrix element on the right-hand side of Eq.~(\ref{qGPD}) can be calculated on the lattice. In Refs.~\cite{Ji:2017oey,Ishikawa:2017faj,Green:2017xeu}, it has been shown that the quark bilinear operator defining $\tilde h$ is multiplicatively renormalized, and the renormalization factor can be calculated nonperturbatively on the lattice. In our previous study of the pion PDF~\cite{Chen:2018fwa}, we chose to calculate the renormalization factor in the RI/MOM scheme, where the counterterm is determined by requiring that it cancels all the loop contributions for the matrix element in an off-shell external quark state at a specific momentum~\cite{Stewart:2017tvs,Chen:2017mzz}.
In other words, the renormalization factor in
\begin{align}
&\tilde h_R(z,P^z, \xi, t, p_z^R, \mu_R)=\nonumber\\
&\hspace*{4em}Z^{-1}(z, p_z^R, \mu_R, a) \tilde h(z,P^z,\xi, t,a)
\end{align}
is fixed by
\begin{align}\label{Zfac}
&Z(z,p^R_z, \mu_R,a)=\left.\frac{{\rm Tr}[\Lambda_{\gamma^t}\cal P]}{{\rm Tr}[\Lambda_{\gamma^t}\cal P]_{\rm tree}}\right|_{\tiny\begin{matrix}p^2=-\mu_R^2 \\ \!\!\!\!p_z=p^R_z\end{matrix}},
\end{align}
where $\Lambda_{\gamma^t}$ is the amputated Green function of the forward quark bilinear operator in Eq.~(\ref{qGPDME}) in an off-shell quark state with momentum $p$. $\cal P$ is a projection operator that defines the RI/MOM renormalization factor, $\mu_R, p_z^R$ are renormalization scales introduced in the RI/MOM scheme. After renormalization, all singular dependence on $a$ has been removed, and $\tilde h_R$ has a well-defined continuum limit. We have suppressed the residual $a$ dependence in $\tilde h_R$. The $Z$ factor defined in Eq.~(\ref{Zfac}) coincides with that of the quark quasi-PDF. Since the UV divergence of the above hadron matrix element depends only on the operator defining it and not on the external state, the same renormalization factor can be used to renormalize the quark quasi-GPD matrix element. After renormalization, ${\tilde h}_R(z,P^z, \xi, t, p_z^R, \mu_R)$ can be converted to $\tilde{H}_q^\pi$ through a Fourier transform
\begin{align}
\tilde{H}_{u-d,R}^\pi(x,\xi,t,P^z,p_z^R, \mu_R) &= \non\\
&\hspace{-7em} \int\frac{d z P^z}{2\pi} e^{i x P^z z}{\tilde h}_R(z,P^z, \xi, t, p_z^R, \mu_R),
\end{align}
which can then be factorized into the normal GPD in the $\overline{\text{MS}}$ scheme convoluted with a perturbative hard matching kernel, up to power corrections that are suppressed by the pion momentum~\cite{Liu:2019urm}
\begin{align} \label{eq:momfact}
&\tilde{H}_{u-d,R}^\pi(x,\xi,t,P^z,p_z^R, \mu_R) \non \\
=& \int_{-1}^1 {dy\over |y|}C\left({x\over y},\frac{\xi}{y},\frac{\mu_{R}}{p_{z}^{R}},\frac{yP^z}{\mu},\frac{yP^z}{p_{z}^{R}}\right)H_{u-d}^\pi(y,\xi,t,\mu)\non\\
+&\mathcal{O}\left({\tilde m_\pi^2\over P_z^2},{\Lambda_\text{QCD}^2\over {P_z^2}}\right),
\end{align}
where $\mu$ is the renormalization scale of the GPD. The matching kernel $C$ has been worked out at one-loop in Ref.~\cite{Liu:2019urm}. At zero skewness $\xi=0$, $C$ is the same as the matching kernel for the PDF that is documented in Refs.~\cite{Chen:2018xof,Liu:2018uuj}.
Ideally, the continuum limit of $\tilde{H}_{u-d,R}$ should be taken before applying the matching so that lattice artifacts can be removed and the rotational symmetry recovered. However, only a single lattice spacing is used in the present work. The continuum limit can be explored in the future once we have more data at different lattice spacings.

For the power corrections, the meson-mass correction associated with the choice of Dirac matrix $\gamma^t$ is identical to that of the helicity distribution worked out in Ref.~\cite{Chen:2016utp} with the replacement $m_\pi^2 \to \tilde m_\pi^2 = m_\pi^2-t/4$. The $\mathcal{O}({\Lambda_\text{QCD}^2/P_z^2})$ correction is parametrically about the same size as the $\mathcal{O}({\tilde m_\pi^2/P_z^2})$ correction (except for very small {or} large $x$ 
{where the correction behaves like $\mathcal{O}({\Lambda_\text{QCD}^2/(x^2(1-x)P_z^2)})$ due to renormalon ambiguity, as argued in Ref.~\cite{Braun:2018brg}}), and is negligible compared with other sources of errors.  

For $\tilde{H}_{u-d}^{\pi^+}$, the matrix element $\tilde h$ is purely real in the isospin symmetric limit which is adopted in this work. {This is because the imaginary part of the matrix element is related to the inverse Fourier transform of $\tilde{H}_{u-d}^{\pi^+}(x)-\tilde{H}_{u-d}^{\pi^+}(-x)$, which is
$\tilde{H}_{u-d}^{\pi^+}(x)+\tilde{H}_{\bar u-\bar d}^{\pi^+}(x) = 0$ after applying the definition of anti-quark distribution $\tilde{H}_{\bar q}(|x|)=-\tilde{H}_{q}(-|x|)$ and the isospin symmetry relation $\tilde{H}^{\pi^+}_{u(d)}=\tilde{H}^{\pi^+}_{\bar d(\bar u)}$.}
Analogously, the real part of the matrix element is related {to the inverse Fourier transform of $\tilde{H}_{u-d}^{\pi^+}(x)+\tilde{H}_{u-d}^{\pi^+}(-x)=\tilde{H}_{u-d}^{\pi^+}(|x|)-\tilde{H}_{\bar u-\bar d}^{\pi^+}(|x|)$, which is the isovector quasi-GPD of the valence quark ($q_{v} \equiv q-\bar q$): $\tilde{H}_{u_v-d_v}^{\pi^+}(|x|)$. }


The above analysis applies not only to the pion quasi-GPD but also to the pion GPD. In the following, we will present our skewless isovector combination of valence quark GPD for the charged pion $\pi^{+}$ as
\begin{equation}
H_{v}^{\pi}(x,t) \equiv \frac{1}{2}H_{u_v-d_v}^{\pi^+}(x,t)=H_{u_v}^{\pi^+}(x,t)=-H_{d_v}^{\pi^+}(x,t) ,
\end{equation}
where the dependence on the renormalization scale $\mu$ is suppressed.

\section{Lattice calculation setup}  
In this work, we use a single ensemble of gauge configurations with $2+1+1$ flavors (degenerate up/down, strange and charm) of highly improved staggered quarks (HISQ)~\cite{Follana:2006rc} generated by the MILC Collaboration~\cite{Bazavov:2012xda}
 with lattice spacing $a \approx 0.12$~fm, pion mass $m_\pi \approx 310$~MeV, and box size $L \approx 3$~fm ($M_\pi L \approx$ 4.5). 
Our calculation is done using clover valence fermions on top of 
one-step hypercubic(HYP)-smeared~\cite{Hasenfratz:2001hp} gauge links, with the clover parameters tuned to recover the lowest pion mass of the staggered quarks in the sea~\cite{Rajan:2017lxk,Bhattacharya:2015wna,Bhattacharya:2015esa,Bhattacharya:2013ehc}. Then we calculate the time-independent, nonlocal (in space, chosen to be in the $z$ direction) correlators of a pion with a finite-$P^z$ boost
\begin{align}
\label{eq:qlat}
& \tilde{h}_\text{lat}(z,P^z,t,a) \\ \nonumber
=&  \frac{1}{2P^0}
  \left\langle \pi^{+}(\vec{P}+\frac{\vec{\Delta}}{2}) \right|
    \bar{q}(z) \Gamma \left( \prod_n U_z(n\hat{z})\right)\tau_{3} q(0)
  \left| \pi^{+}(\vec{P}-\frac{\vec{\Delta}}{2}) \right\rangle,
\end{align}
where $U_z$ is a discrete gauge link in the $z$ direction, $\vec{P}=\{0,0,P^z\}$ is the momentum of the pion, $\Gamma=\gamma^t$ and $\vec{\Delta}$ is the momentum transfer between initial and final pion. 
In this work, we only deal with the zero-skewness limit $\xi=0$, where the matching coincides with that for the PDF.
We use 3 boosted pion momenta, $P^z=\{0,0, n \frac{2\pi}{L}\}$ 
with $n \in \{2,3,4\}$, which correspond to 0.86, 1.32 and 1.74~GeV, respectively.
The initial and final pion momenta are obtained from $\{0,0, n \frac{2\pi}{L}\} \mp \vec{\Delta}/2$, where $\vec{\Delta} = \{n_x,n_y,0\}\frac{2\pi}{L}$ with $n_x^2+n_y^2 \leq 5$ f{and $P_{0}=\sqrt{(P^{z})^{2}-t/4+m_{\pi}^{2}}$}. 
We carefully tune the Gaussian smearing parameter to best control the excited state and use four source-sink separations, 0.72, 0.84, 0.96 and 1.08 fm to help us remove excited-state contamination from our three-point correlators fits to extract the pion matrix element. We use 1840 configurations with total measurements of 29440, 29440, 58880 and 58880 from the smallest source-sink separation to largest one. After we obtain the pion form factors at each momentum, we momentum-average the spatial symmetry in the $x-y$ plane for each momentum transfer $t$.

To make sure that we have full control of excited-state contamination, we analyze our data using different source-sink separations and using different levels of excited-state treatment. 
First, we use the ``two-sim'' analysis described in Ref.~\cite{Bhattacharya:2013ehc} to obtain the ground-state pion matrix elements using all 4 source-sink separations. 
The ``two-sim'' analysis only takes account of the leading excited-state contamination coming from the excited- and ground-state mixing. This is the same level as other commonly used methods, such as the ``summation'' method~\cite{Capitani:2012gj}.
A second extraction uses only the largest two separations; if there is a significant excited-state contamination at the smaller source-sink separation, 0.72 and 0.84~fm, we should see inconsistency in the ground-state matrix element between this analysis and earlier ones. 
Finally, we use the ``two-twoRR'' analysis (see Ref.~\cite{Bhattacharya:2013ehc} for details), which includes an additional matrix element related to excited states. Given the same input source-sink separations for the three-point correlators, the extracted ground-state matrix elements should be noisier, since more fit parameters are used. 
\footnote{The detailed procedure for treating excited-state systematics can be found in Ref.~\cite{Bhattacharya:2013ehc} for the nucleon-charge case.}
All the above analyses generate consistent ground-state nucleon matrix elements. {A few example fit plots from a subset of data are shown in Fig.~\ref{fig:ME-ratios}. In the GPD analysis to be presented below, we use matrix elements from the ``two-twoRR'' analysis.}

\begin{figure}[t]
\includegraphics[width=.5\textwidth]{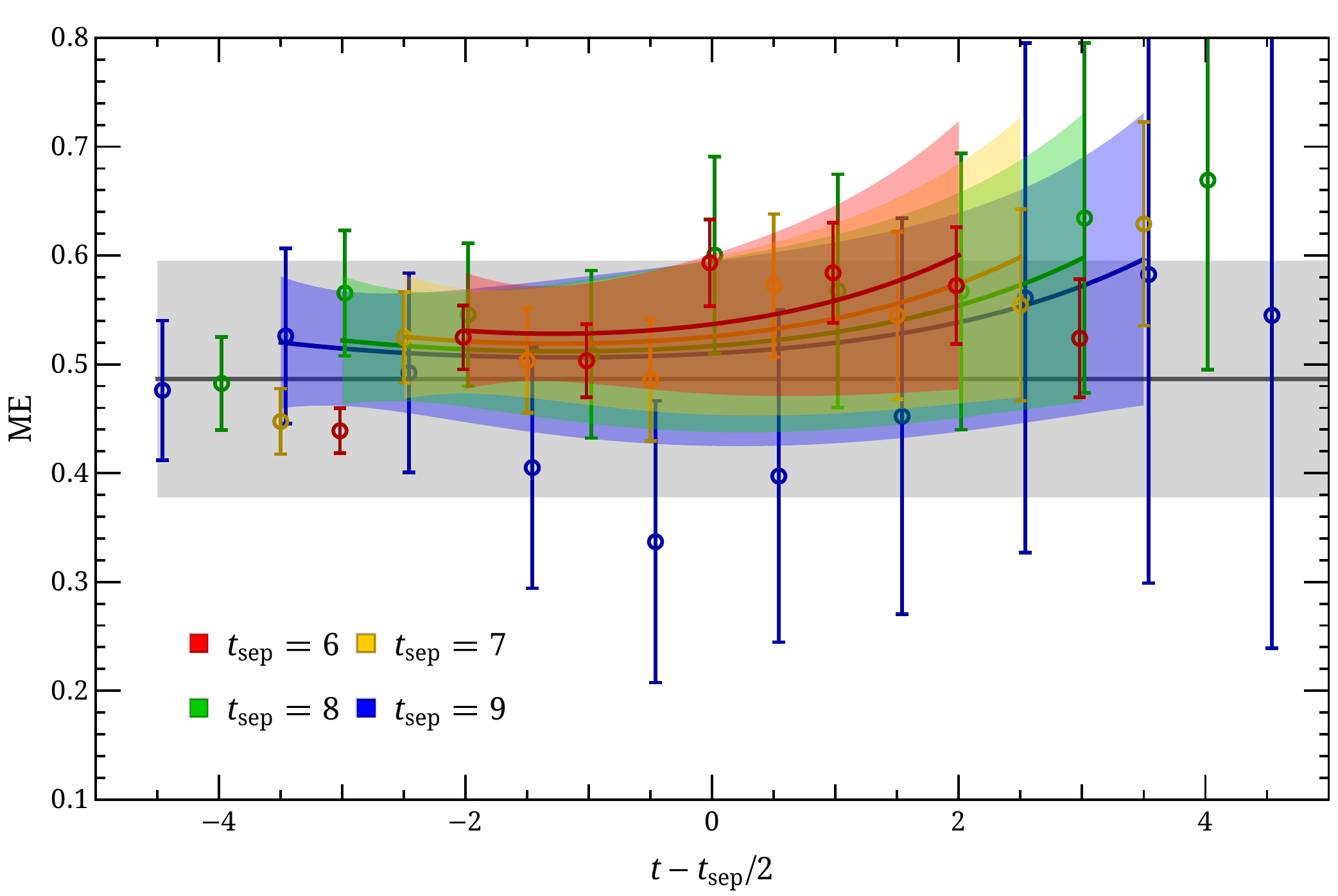}
\includegraphics[width=.5\textwidth]{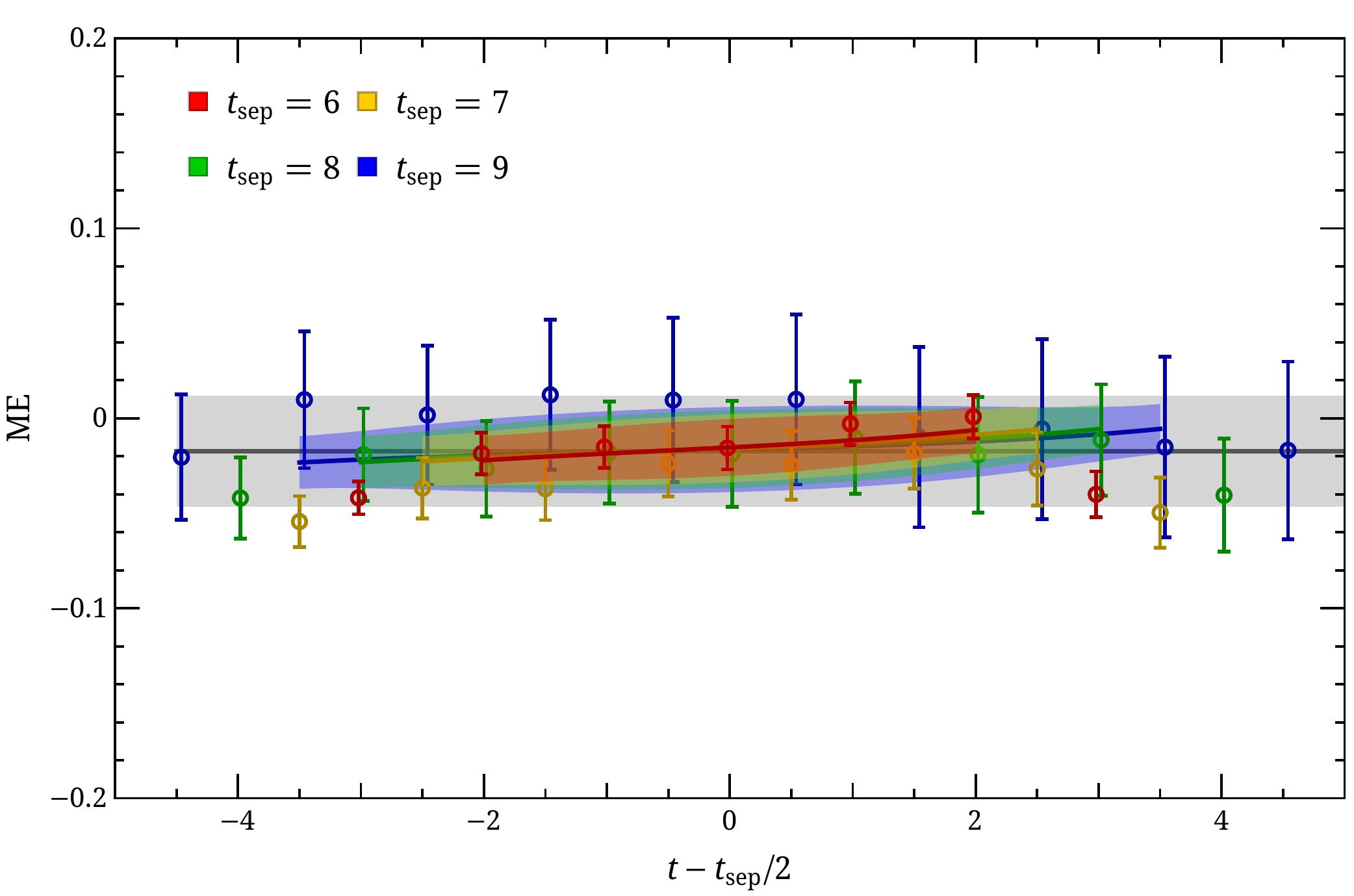}
\caption{{Example three-point ratio plots  as functions of the operator insertion time $t$ from the real pion matrix elements for $z=-3$ with momentum transfer of $\{2,1,0\}\times(2\pi/L)$ (top) and for $z=6$ with momentum transfer of $\{1,0,0\}\times(2\pi/L)$ (bottom). The red, yellow, green, and blue bands are the reconstructed ratios from the fits to source-sink separation $t_\text{sep}=6,7,8,9$, respectively, and the gray band shows the ground-state matrix elements from the ``two-simRR'' fits. }} 
\label{fig:ME-ratios}
\end{figure}

Fig.~\ref{fig:ME-tsep} shows the real part of the bare and RI/MOM renormalized matrix elements for $P^z\in\{4\pi/L, 8\pi/L\}\approx \{0.87, 1.74\}$~GeV, and $t\in\{0,-2,-5\}\times(2 \pi/L)^{2}$.  The renormalization scales in the RI/MOM parameters have been chosen as $\mu_R=3.7$~GeV, $p_z^R=6\times 2\pi/L$. {The error bars include statistical errors and the errors from the excited-state contamination. Systematic errors from renormalization scale dependence associated with one-loop matching, lattice spacing, non-physical pion mass and so on are not included in this figure.  }
For $z=0$,  the error at $t=0$ is much smaller than $t \ne 0$, due to charge conservation. The matrix elements at $z=0$, which are not changed by the renormalization, are the values of the pion isovector form factor. It is decreasing in $|t|$ as expected. {Our form factor also agrees with 
the previous result (shown as points at $z=0$ in Fig.~\ref{fig:ME-tsep}) obtained in Ref.~\cite{Brommel:2006ww}, which were determined from a fitted form to lattice data with a wide range of pion masses and lattice spacings and setting the pion mass to the same value used here. The errors were estimated from the difference between two fitting forms used in Ref.~\cite{Brommel:2006ww}. The error of our matrix elements at $z=0$ is larger than that in the form factor calculation, mainly because the latter is equivalent to having $P^z=0$ while we need nonzero $P^z$ to access the full distribution}.

\section{Numerical results and discussion}  

Now we present our numerical results for the valence quark GPD in the pion. As mentioned previously, the bare quasi-GPD matrix element calculated on the lattice can be renormalized by the RI/MOM renormalization factors for the quasi-PDF matrix element, which have been computed in Ref.~\cite{Liu:2018uuj}. The momentum distribution is then given by 
\begin{equation}\label{FThR}
\tilde H_v^\pi(x)=\int \frac{dz}{4\pi}e^{ix z P^z}\tilde h_{\text{lat},R}(z),
\end{equation}
where only the $z$ and $x$ dependence is shown for simplicity. Following our earlier work~\cite{Lin:2017ani} on the nucleon PDF, we also apply the ``derivative'' method~\cite{Lin:2017ani} to improve the truncation error in the Fourier transform in Eq.~(\ref{FThR}).
We then apply the one-loop matching and meson-mass corrections, where the latter turn out to be numerically negligible. 


\begin{figure}[tbp]
\includegraphics[width=.45\textwidth]{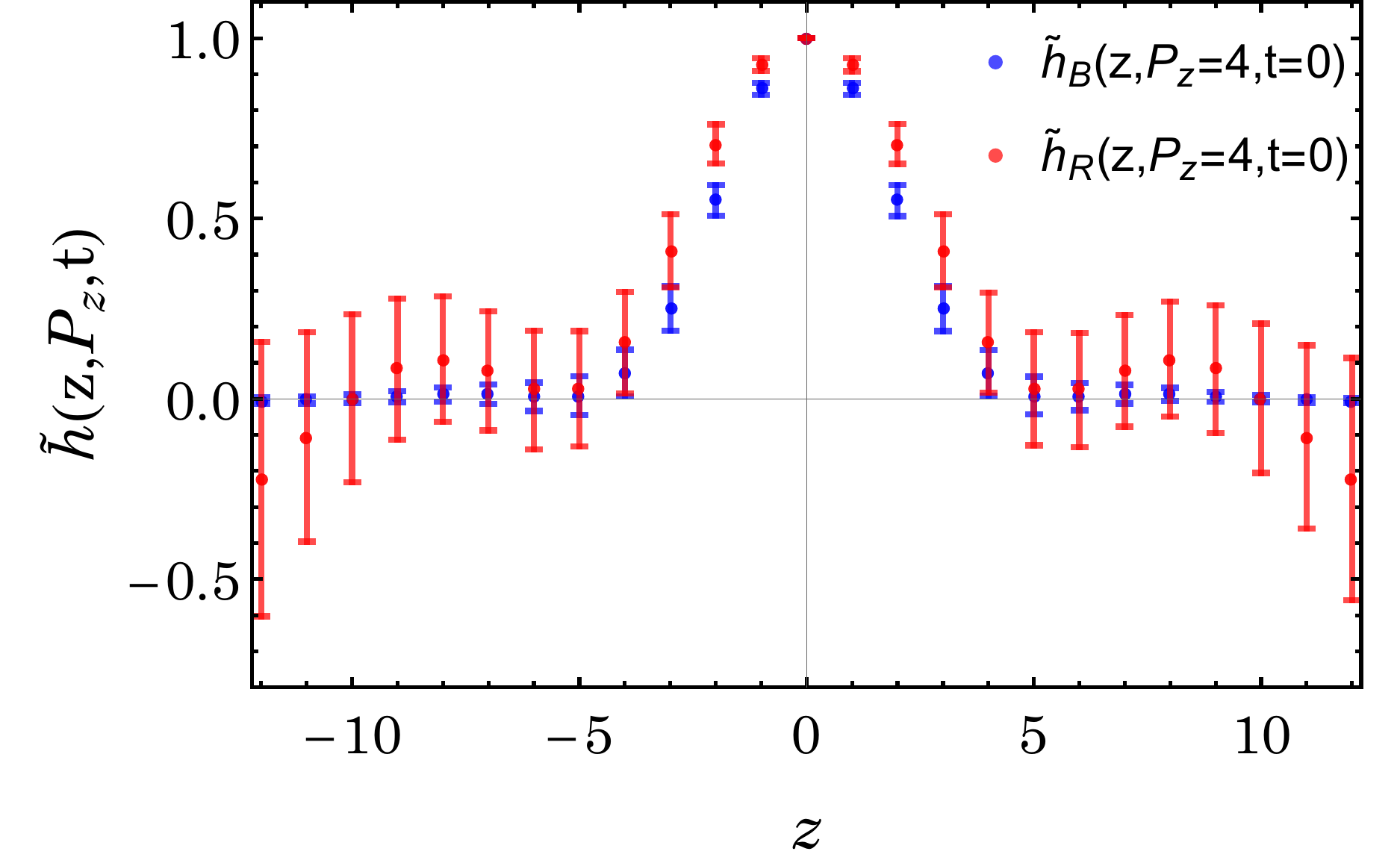}
\includegraphics[width=.45\textwidth]{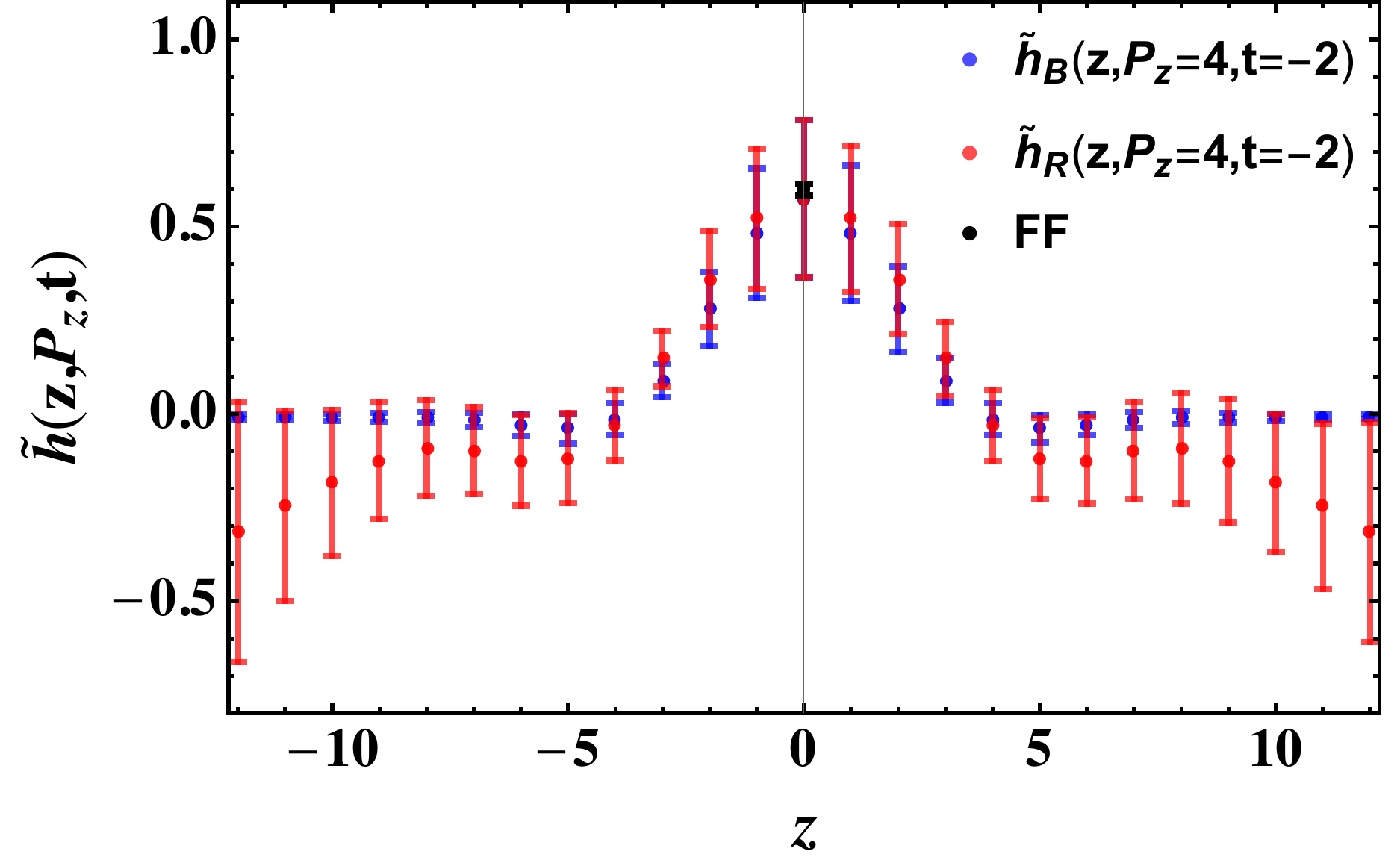}
\includegraphics[width=.45\textwidth]{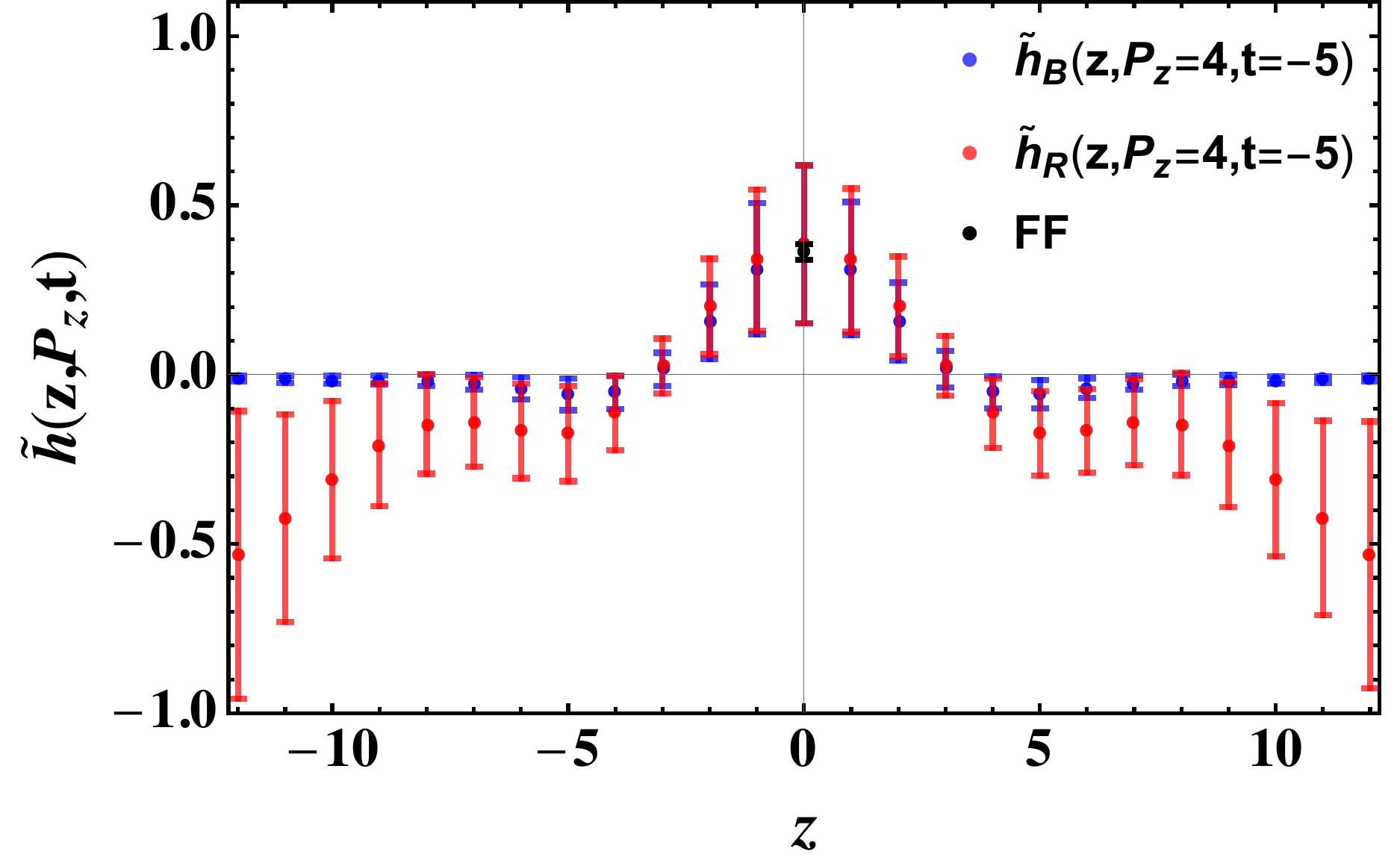}
\caption{The real part of the bare and renormalized matrix elements, ${\tilde h}_B(z, P^z, t)$ and ${\tilde h}_R(z, P^z, t)$, for the zero-skewness isovector valence quark GPD of the pion. The averaged pion momentum in the $z$-direction is $P^z=4$ (left to right, in units of $(2\pi/L)$) and the momentum transfer squared is $t=\{0,-2,-5\}$ (top to bottom, in units of $(2\pi/L)^{2}$). {Also plotted at $z=0$ 
are the form factors (FF) from previous lattice calculations~\cite{Brommel:2006ww} at the same pion mass.}}  
\label{fig:ME-tsep}
\end{figure}


\begin{figure}[t]
\includegraphics[width=.5\textwidth]{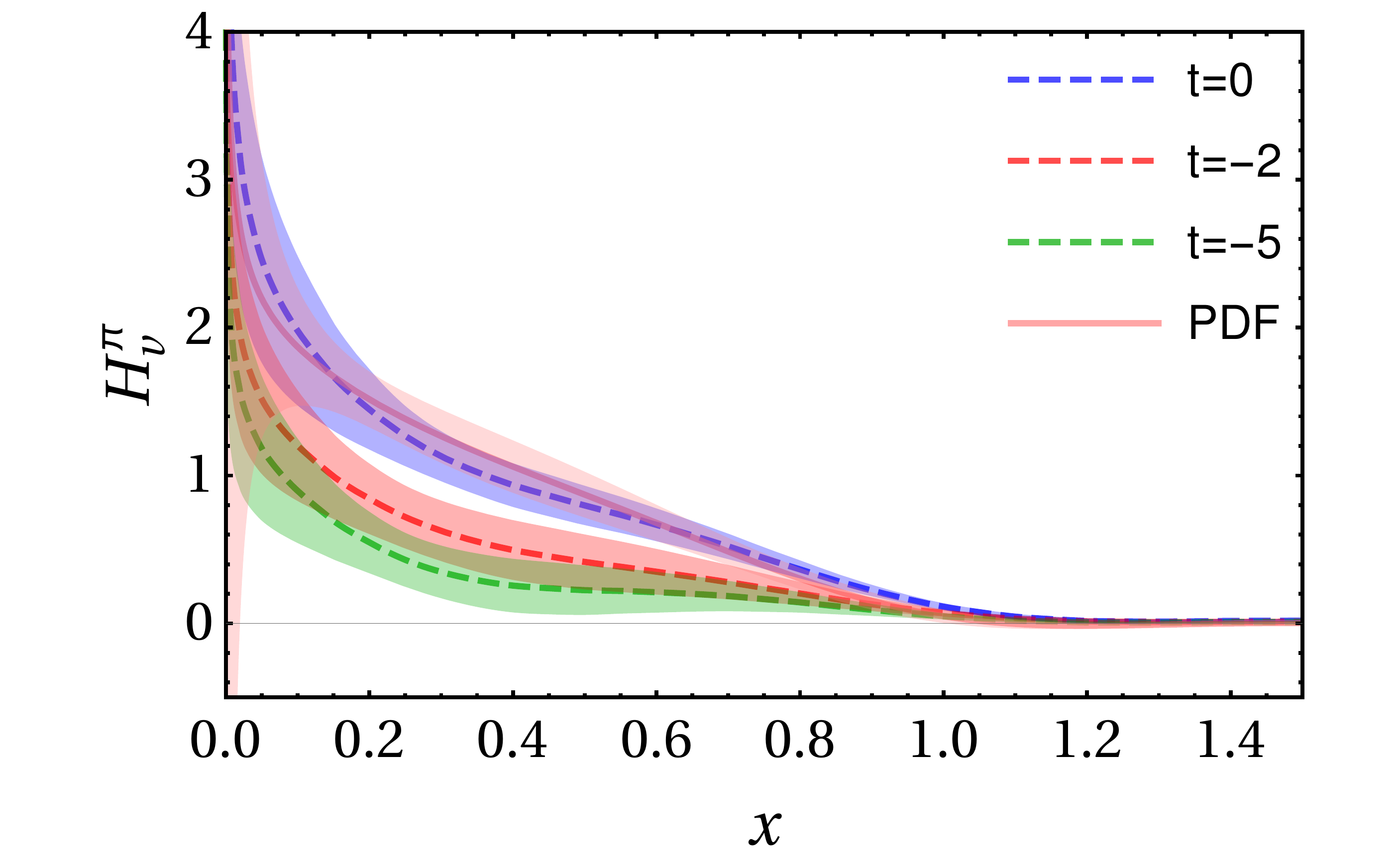}
\caption{
The zero-skewness pion valence quark GPD {$H^{\pi^{+}}_{v}(x,\xi=0,t,\mu=4\text{GeV})$
for $t=\{0, -2, -5\}(2\pi/L)^2$} after one-loop matching  and the meson-mass correction. {``PDF" denotes the pion PDF result in Ref.~\cite{Chen:2018fwa}}.}
\label{fig:xqvpi}
\end{figure}

In Fig.~\ref{fig:xqvpi}, we show the results of the valence-quark distribution $H_{v}^{\pi}(x,t,\mu)$ for different values of $t$ with the renormalization scale $\mu=4$ GeV. We have inverted the factorization formula in Eq.~(\ref{eq:momfact}) by perturbatively expanding the matching kernel $C$ to ${\cal O}(\alpha_s)$. Also, the meson-mass power corrections have been removed. For the RI/MOM renormalization of the quark quasi-GPD, we have chosen $\mu_R=3.7$~GeV, $p_z^R=6\times 2\pi/L$. The error bands in Fig.~\ref{fig:xqvpi} include statistical as well as systematic errors of $p_z^R$ dependence by varying it between $4\times 2\pi/L$ and $8\times 2\pi/L$. The curve with $t=0$ is consistent with our previous 
result~\cite{Chen:2018fwa} within errors. {As a consistency check, we also tested that the $x$-integral of the GPDs in Fig.~\ref{fig:xqvpi} reproduces the form factors in Fig.~\ref{fig:ME-tsep} within 1 standard deviation.} The impact parameter-space distribution can in principle be obtained by Fourier transforming the $t$ dependence to the impact parameter $b_{\perp}$ dependence. However, we have only the results for very few values of $t$ in this work.

For the kinematic dependence of $H_{v}^{\pi}(x,t,\mu)$, a naive functional form is
\begin{equation}\label{factform}
H_{v}^{\pi}(x,t,\mu)=q_v^\pi(x,\mu)F_{u-d}^\pi(t),
\end{equation}
where $q^\pi_{v}$ is the pion valence-quark PDF satisfying $\int dx\, q^\pi_{v}(x,\mu) = 1$, and $F^\pi_{u-d}(t)$ is the isovector form factor of $\pi^{+}$ with the normalization $F^\pi_{u-d}(0)=1$. { This parametrization is simple, but not favored by the study of the GPD asymptotic behavior at $x\to 1$~\cite{Burkardt:2002hr,Diehl:2004cx}. On the lattice side, Eq.(\ref{factform}) implies factorization of the bare matrix element
\begin{equation}\label{factorization1}
\tilde{h}_\text{lat}(z,P^z,t,a)= F^\pi_{u-d}(t) \tilde{h}_\text{lat}(z,P^z,t\emph{}=0,a),
\end{equation} 
which makes it easier to be checked with lattice QCD. 
} 

Another useful parametrization is 
\begin{equation}\label{fitform}
H_{v}^{\pi}(x,t,\mu)=q_v^\pi(x,\mu)\exp[t f_q(x,\mu)],
\end{equation}
{with $f_q$ an unknown function.} This parametrization 
has been used to fit the unpolarized zero-skewness GPD of the nucleon and in discussing the fit to experimental data for the nucleon form factors~\cite{Diehl:2004cx}. 

{With current uncertainties, our results are consistent with both parametrizations. Future high-statistics studies will be able to provide guidance to the kinematic dependence of the GPDs, and in particular allow differentiation between various models that are commonly used.}

\section{Summary} 

We have presented the first lattice calculation of the valence-quark generalized parton distribution of the pion using the LaMET approach. We have focused on the zero-skewness limit, where the GPD has a probability-density interpretation in the longitudinal Bjorken $x$ and the transverse impact-parameter distributions. 
Our calculation is done using clover valence fermions on an ensemble of gauge configurations with $2+1+1$ flavors (degenerate up/down, strange and charm) of highly improved staggered quarks (HISQ) with lattice spacing $a \approx 0.12$~fm, box size $L \approx 3$~fm and pion mass $m_\pi \approx 310$~MeV. The parton distribution function and form factor are reproduced as special limits of the GPD as expected. 
{Future studies using higher-statistics data will be crucial to provide guidance to the kinematic dependence of the GPDs and to differentiate models that are commonly used.}
\vspace{1em}

\section*{Acknowledgments}
We thank the MILC Collaboration for sharing the lattices used to perform this study. The LQCD calculations were performed using the Chroma software suite~\cite{Edwards:2004sx}. 
This research used resources of the National Energy Research Scientific Computing Center, a DOE Office of Science User Facility supported by the Office of Science of the U.S. Department of Energy under Contract No. DE-AC02-05CH11231 through ALCC and ERCAP; facilities of the USQCD Collaboration, which are funded by the Office of Science of the U.S. Department of Energy, and supported in part by Michigan State University through computational resources provided by the Institute for Cyber-Enabled Research. JWC is partly supported by the Ministry of Science and Technology, Taiwan, under Grant No. 105-2112-M-002-017-MY3 and the Kenda Foundation. 
HL is supported by the US National Science Foundation under grant PHY 1653405 ``CAREER: Constraining Parton Distribution Functions for New-Physics Searches''. JHZ thanks M. Diehl and A. Sch\"afer for helpful discussions. He is supported by the SFB/TRR-55 grant ``Hadron Physics from Lattice QCD''.

\bibliographystyle{apsrev4-1}
\bibliography{refs}

\begin{thebibliography}{98}%
\makeatletter
\providecommand \@ifxundefined [1]{%
 \@ifx{#1\undefined}
}%
\providecommand \@ifnum [1]{%
 \ifnum #1\expandafter \@firstoftwo
 \else \expandafter \@secondoftwo
 \fi
}%
\providecommand \@ifx [1]{%
 \ifx #1\expandafter \@firstoftwo
 \else \expandafter \@secondoftwo
 \fi
}%
\providecommand \natexlab [1]{#1}%
\providecommand \enquote  [1]{``#1''}%
\providecommand \bibnamefont  [1]{#1}%
\providecommand \bibfnamefont [1]{#1}%
\providecommand \citenamefont [1]{#1}%
\providecommand \href@noop [0]{\@secondoftwo}%
\providecommand \href [0]{\begingroup \@sanitize@url \@href}%
\providecommand \@href[1]{\@@startlink{#1}\@@href}%
\providecommand \@@href[1]{\endgroup#1\@@endlink}%
\providecommand \@sanitize@url [0]{\catcode `\\12\catcode `\$12\catcode
  `\&12\catcode `\#12\catcode `\^12\catcode `\_12\catcode `\%12\relax}%
\providecommand \@@startlink[1]{}%
\providecommand \@@endlink[0]{}%
\providecommand \url  [0]{\begingroup\@sanitize@url \@url }%
\providecommand \@url [1]{\endgroup\@href {#1}{\urlprefix }}%
\providecommand \urlprefix  [0]{URL }%
\providecommand \Eprint [0]{\href }%
\providecommand \doibase [0]{http://dx.doi.org/}%
\providecommand \selectlanguage [0]{\@gobble}%
\providecommand \bibinfo  [0]{\@secondoftwo}%
\providecommand \bibfield  [0]{\@secondoftwo}%
\providecommand \translation [1]{[#1]}%
\providecommand \BibitemOpen [0]{}%
\providecommand \bibitemStop [0]{}%
\providecommand \bibitemNoStop [0]{.\EOS\space}%
\providecommand \EOS [0]{\spacefactor3000\relax}%
\providecommand \BibitemShut  [1]{\csname bibitem#1\endcsname}%
\let\auto@bib@innerbib\@empty
\bibitem [{\citenamefont {Ji}(1997{\natexlab{a}})}]{Ji:1996ek}%
  \BibitemOpen
  \bibfield  {author} {\bibinfo {author} {\bibfnamefont {X.-D.}\ \bibnamefont
  {Ji}},\ }\href {\doibase 10.1103/PhysRevLett.78.610} {\bibfield  {journal}
  {\bibinfo  {journal} {Phys. Rev. Lett.}\ }\textbf {\bibinfo {volume} {78}},\
  \bibinfo {pages} {610} (\bibinfo {year} {1997}{\natexlab{a}})},\ \Eprint
  {http://arxiv.org/abs/hep-ph/9603249} {arXiv:hep-ph/9603249 [hep-ph]}
  \BibitemShut {NoStop}%
\bibitem [{\citenamefont {Ji}(1997{\natexlab{b}})}]{Ji:1996nm}%
  \BibitemOpen
  \bibfield  {author} {\bibinfo {author} {\bibfnamefont {X.-D.}\ \bibnamefont
  {Ji}},\ }\href {\doibase 10.1103/PhysRevD.55.7114} {\bibfield  {journal}
  {\bibinfo  {journal} {Phys. Rev.}\ }\textbf {\bibinfo {volume} {D55}},\
  \bibinfo {pages} {7114} (\bibinfo {year} {1997}{\natexlab{b}})},\ \Eprint
  {http://arxiv.org/abs/hep-ph/9609381} {arXiv:hep-ph/9609381 [hep-ph]}
  \BibitemShut {NoStop}%
\bibitem [{\citenamefont {Müller}\ \emph {et~al.}(1994)\citenamefont
  {Müller}, \citenamefont {Robaschik}, \citenamefont {Geyer}, \citenamefont
  {Dittes},\ and\ \citenamefont {Hořejši}}]{Mueller:1998fv}%
  \BibitemOpen
  \bibfield  {author} {\bibinfo {author} {\bibfnamefont {D.}~\bibnamefont
  {Müller}}, \bibinfo {author} {\bibfnamefont {D.}~\bibnamefont {Robaschik}},
  \bibinfo {author} {\bibfnamefont {B.}~\bibnamefont {Geyer}}, \bibinfo
  {author} {\bibfnamefont {F.~M.}\ \bibnamefont {Dittes}}, \ and\ \bibinfo
  {author} {\bibfnamefont {J.}~\bibnamefont {Hořejši}},\ }\href {\doibase
  10.1002/prop.2190420202} {\bibfield  {journal} {\bibinfo  {journal} {Fortsch.
  Phys.}\ }\textbf {\bibinfo {volume} {42}},\ \bibinfo {pages} {101} (\bibinfo
  {year} {1994})},\ \Eprint {http://arxiv.org/abs/hep-ph/9812448}
  {arXiv:hep-ph/9812448 [hep-ph]} \BibitemShut {NoStop}%
\bibitem [{\citenamefont {Airapetian}\ \emph {et~al.}(2001)\citenamefont
  {Airapetian} \emph {et~al.}}]{Airapetian:2001yk}%
  \BibitemOpen
  \bibfield  {author} {\bibinfo {author} {\bibfnamefont {A.}~\bibnamefont
  {Airapetian}} \emph {et~al.} (\bibinfo {collaboration} {HERMES}),\ }\href
  {\doibase 10.1103/PhysRevLett.87.182001} {\bibfield  {journal} {\bibinfo
  {journal} {Phys. Rev. Lett.}\ }\textbf {\bibinfo {volume} {87}},\ \bibinfo
  {pages} {182001} (\bibinfo {year} {2001})},\ \Eprint
  {http://arxiv.org/abs/hep-ex/0106068} {arXiv:hep-ex/0106068 [hep-ex]}
  \BibitemShut {NoStop}%
\bibitem [{\citenamefont {Adloff}\ \emph {et~al.}(2001)\citenamefont {Adloff}
  \emph {et~al.}}]{Adloff:2001cn}%
  \BibitemOpen
  \bibfield  {author} {\bibinfo {author} {\bibfnamefont {C.}~\bibnamefont
  {Adloff}} \emph {et~al.} (\bibinfo {collaboration} {H1}),\ }\href {\doibase
  10.1016/S0370-2693(01)00939-X} {\bibfield  {journal} {\bibinfo  {journal}
  {Phys. Lett.}\ }\textbf {\bibinfo {volume} {B517}},\ \bibinfo {pages} {47}
  (\bibinfo {year} {2001})},\ \Eprint {http://arxiv.org/abs/hep-ex/0107005}
  {arXiv:hep-ex/0107005 [hep-ex]} \BibitemShut {NoStop}%
\bibitem [{\citenamefont {Chekanov}\ \emph {et~al.}(2003)\citenamefont
  {Chekanov} \emph {et~al.}}]{Chekanov:2003ya}%
  \BibitemOpen
  \bibfield  {author} {\bibinfo {author} {\bibfnamefont {S.}~\bibnamefont
  {Chekanov}} \emph {et~al.} (\bibinfo {collaboration} {ZEUS}),\ }\href
  {\doibase 10.1016/j.physletb.2003.08.048} {\bibfield  {journal} {\bibinfo
  {journal} {Phys. Lett.}\ }\textbf {\bibinfo {volume} {B573}},\ \bibinfo
  {pages} {46} (\bibinfo {year} {2003})},\ \Eprint
  {http://arxiv.org/abs/hep-ex/0305028} {arXiv:hep-ex/0305028 [hep-ex]}
  \BibitemShut {NoStop}%
\bibitem [{\citenamefont {Defurne}\ \emph {et~al.}(2015)\citenamefont {Defurne}
  \emph {et~al.}}]{Defurne:2015kxq}%
  \BibitemOpen
  \bibfield  {author} {\bibinfo {author} {\bibfnamefont {M.}~\bibnamefont
  {Defurne}} \emph {et~al.} (\bibinfo {collaboration} {Jefferson Lab Hall A}),\
  }\href {\doibase 10.1103/PhysRevC.92.055202} {\bibfield  {journal} {\bibinfo
  {journal} {Phys. Rev.}\ }\textbf {\bibinfo {volume} {C92}},\ \bibinfo {pages}
  {055202} (\bibinfo {year} {2015})},\ \Eprint
  {http://arxiv.org/abs/1504.05453} {arXiv:1504.05453 [nucl-ex]} \BibitemShut
  {NoStop}%
\bibitem [{\citenamefont {Jo}\ \emph {et~al.}(2015)\citenamefont {Jo} \emph
  {et~al.}}]{Jo:2015ema}%
  \BibitemOpen
  \bibfield  {author} {\bibinfo {author} {\bibfnamefont {H.~S.}\ \bibnamefont
  {Jo}} \emph {et~al.} (\bibinfo {collaboration} {CLAS}),\ }\href {\doibase
  10.1103/PhysRevLett.115.212003} {\bibfield  {journal} {\bibinfo  {journal}
  {Phys. Rev. Lett.}\ }\textbf {\bibinfo {volume} {115}},\ \bibinfo {pages}
  {212003} (\bibinfo {year} {2015})},\ \Eprint
  {http://arxiv.org/abs/1504.02009} {arXiv:1504.02009 [hep-ex]} \BibitemShut
  {NoStop}%
\bibitem [{\citenamefont {Seder}\ \emph {et~al.}(2015)\citenamefont {Seder}
  \emph {et~al.}}]{Seder:2014cdc}%
  \BibitemOpen
  \bibfield  {author} {\bibinfo {author} {\bibfnamefont {E.}~\bibnamefont
  {Seder}} \emph {et~al.} (\bibinfo {collaboration} {CLAS}),\ }\href {\doibase
  10.1103/PhysRevLett.114.089901, 10.1103/PhysRevLett.114.032001} {\bibfield
  {journal} {\bibinfo  {journal} {Phys. Rev. Lett.}\ }\textbf {\bibinfo
  {volume} {114}},\ \bibinfo {pages} {032001} (\bibinfo {year} {2015})},\
  \bibinfo {note} {[Addendum: Phys. Rev. Lett.114,no.8,089901(2015)]},\ \Eprint
  {http://arxiv.org/abs/1410.6615} {arXiv:1410.6615 [hep-ex]} \BibitemShut
  {NoStop}%
\bibitem [{\citenamefont {Dudek}\ \emph {et~al.}(2012)\citenamefont {Dudek}
  \emph {et~al.}}]{Dudek:2012vr}%
  \BibitemOpen
  \bibfield  {author} {\bibinfo {author} {\bibfnamefont {J.}~\bibnamefont
  {Dudek}} \emph {et~al.},\ }\href {\doibase 10.1140/epja/i2012-12187-1}
  {\bibfield  {journal} {\bibinfo  {journal} {Eur. Phys. J.}\ }\textbf
  {\bibinfo {volume} {A48}},\ \bibinfo {pages} {187} (\bibinfo {year}
  {2012})},\ \Eprint {http://arxiv.org/abs/1208.1244} {arXiv:1208.1244
  [hep-ex]} \BibitemShut {NoStop}%
\bibitem [{\citenamefont {Belitsky}\ and\ \citenamefont
  {Radyushkin}(2005)}]{Belitsky:2005qn}%
  \BibitemOpen
  \bibfield  {author} {\bibinfo {author} {\bibfnamefont {A.~V.}\ \bibnamefont
  {Belitsky}}\ and\ \bibinfo {author} {\bibfnamefont {A.~V.}\ \bibnamefont
  {Radyushkin}},\ }\href {\doibase 10.1016/j.physrep.2005.06.002} {\bibfield
  {journal} {\bibinfo  {journal} {Phys. Rept.}\ }\textbf {\bibinfo {volume}
  {418}},\ \bibinfo {pages} {1} (\bibinfo {year} {2005})},\ \Eprint
  {http://arxiv.org/abs/hep-ph/0504030} {arXiv:hep-ph/0504030 [hep-ph]}
  \BibitemShut {NoStop}%
\bibitem [{\citenamefont {Gockeler}\ \emph {et~al.}(2004)\citenamefont
  {Gockeler}, \citenamefont {Horsley}, \citenamefont {Pleiter}, \citenamefont
  {Rakow}, \citenamefont {Schafer}, \citenamefont {Schierholz},\ and\
  \citenamefont {Schroers}}]{Gockeler:2003jfa}%
  \BibitemOpen
  \bibfield  {author} {\bibinfo {author} {\bibfnamefont {M.}~\bibnamefont
  {Gockeler}}, \bibinfo {author} {\bibfnamefont {R.}~\bibnamefont {Horsley}},
  \bibinfo {author} {\bibfnamefont {D.}~\bibnamefont {Pleiter}}, \bibinfo
  {author} {\bibfnamefont {P.~E.~L.}\ \bibnamefont {Rakow}}, \bibinfo {author}
  {\bibfnamefont {A.}~\bibnamefont {Schafer}}, \bibinfo {author} {\bibfnamefont
  {G.}~\bibnamefont {Schierholz}}, \ and\ \bibinfo {author} {\bibfnamefont
  {W.}~\bibnamefont {Schroers}} (\bibinfo {collaboration} {QCDSF}),\ }\href
  {\doibase 10.1103/PhysRevLett.92.042002} {\bibfield  {journal} {\bibinfo
  {journal} {Phys. Rev. Lett.}\ }\textbf {\bibinfo {volume} {92}},\ \bibinfo
  {pages} {042002} (\bibinfo {year} {2004})},\ \Eprint
  {http://arxiv.org/abs/hep-ph/0304249} {arXiv:hep-ph/0304249 [hep-ph]}
  \BibitemShut {NoStop}%
\bibitem [{\citenamefont {Hagler}\ \emph {et~al.}(2003)\citenamefont {Hagler},
  \citenamefont {Negele}, \citenamefont {Renner}, \citenamefont {Schroers},
  \citenamefont {Lippert},\ and\ \citenamefont {Schilling}}]{Hagler:2003jd}%
  \BibitemOpen
  \bibfield  {author} {\bibinfo {author} {\bibfnamefont {P.}~\bibnamefont
  {Hagler}}, \bibinfo {author} {\bibfnamefont {J.~W.}\ \bibnamefont {Negele}},
  \bibinfo {author} {\bibfnamefont {D.~B.}\ \bibnamefont {Renner}}, \bibinfo
  {author} {\bibfnamefont {W.}~\bibnamefont {Schroers}}, \bibinfo {author}
  {\bibfnamefont {T.}~\bibnamefont {Lippert}}, \ and\ \bibinfo {author}
  {\bibfnamefont {K.}~\bibnamefont {Schilling}} (\bibinfo {collaboration}
  {LHPC, SESAM}),\ }\href {\doibase 10.1103/PhysRevD.68.034505} {\bibfield
  {journal} {\bibinfo  {journal} {Phys. Rev.}\ }\textbf {\bibinfo {volume}
  {D68}},\ \bibinfo {pages} {034505} (\bibinfo {year} {2003})},\ \Eprint
  {http://arxiv.org/abs/hep-lat/0304018} {arXiv:hep-lat/0304018 [hep-lat]}
  \BibitemShut {NoStop}%
\bibitem [{\citenamefont {LHPC}\ \emph {et~al.}(2004)\citenamefont {LHPC},
  \citenamefont {Hagler}, \citenamefont {Negele}, \citenamefont {Renner},
  \citenamefont {Schroers}, \citenamefont {Lippert},\ and\ \citenamefont
  {Schilling}}]{LHPC:2003aa}%
  \BibitemOpen
  \bibfield  {author} {\bibinfo {author} {\bibnamefont {LHPC}}, \bibinfo
  {author} {\bibfnamefont {P.}~\bibnamefont {Hagler}}, \bibinfo {author}
  {\bibfnamefont {J.~W.}\ \bibnamefont {Negele}}, \bibinfo {author}
  {\bibfnamefont {D.~B.}\ \bibnamefont {Renner}}, \bibinfo {author}
  {\bibfnamefont {W.}~\bibnamefont {Schroers}}, \bibinfo {author}
  {\bibfnamefont {T.}~\bibnamefont {Lippert}}, \ and\ \bibinfo {author}
  {\bibfnamefont {K.}~\bibnamefont {Schilling}} (\bibinfo {collaboration}
  {LHPC, SESAM}),\ }\href {\doibase 10.1103/PhysRevLett.93.112001} {\bibfield
  {journal} {\bibinfo  {journal} {Phys. Rev. Lett.}\ }\textbf {\bibinfo
  {volume} {93}},\ \bibinfo {pages} {112001} (\bibinfo {year} {2004})},\
  \Eprint {http://arxiv.org/abs/hep-lat/0312014} {arXiv:hep-lat/0312014
  [hep-lat]} \BibitemShut {NoStop}%
\bibitem [{\citenamefont {Gockeler}\ \emph {et~al.}(2005)\citenamefont
  {Gockeler}, \citenamefont {Hagler}, \citenamefont {Horsley}, \citenamefont
  {Pleiter}, \citenamefont {Rakow}, \citenamefont {Schafer}, \citenamefont
  {Schierholz},\ and\ \citenamefont {Zanotti}}]{Gockeler:2005cj}%
  \BibitemOpen
  \bibfield  {author} {\bibinfo {author} {\bibfnamefont {M.}~\bibnamefont
  {Gockeler}}, \bibinfo {author} {\bibfnamefont {P.}~\bibnamefont {Hagler}},
  \bibinfo {author} {\bibfnamefont {R.}~\bibnamefont {Horsley}}, \bibinfo
  {author} {\bibfnamefont {D.}~\bibnamefont {Pleiter}}, \bibinfo {author}
  {\bibfnamefont {P.~E.~L.}\ \bibnamefont {Rakow}}, \bibinfo {author}
  {\bibfnamefont {A.}~\bibnamefont {Schafer}}, \bibinfo {author} {\bibfnamefont
  {G.}~\bibnamefont {Schierholz}}, \ and\ \bibinfo {author} {\bibfnamefont
  {J.~M.}\ \bibnamefont {Zanotti}} (\bibinfo {collaboration} {QCDSF, UKQCD}),\
  }\href {\doibase 10.1016/j.physletb.2005.09.002} {\bibfield  {journal}
  {\bibinfo  {journal} {Phys. Lett.}\ }\textbf {\bibinfo {volume} {B627}},\
  \bibinfo {pages} {113} (\bibinfo {year} {2005})},\ \Eprint
  {http://arxiv.org/abs/hep-lat/0507001} {arXiv:hep-lat/0507001 [hep-lat]}
  \BibitemShut {NoStop}%
\bibitem [{\citenamefont {Hagler}(2010)}]{Hagler:2009ni}%
  \BibitemOpen
  \bibfield  {author} {\bibinfo {author} {\bibfnamefont {P.}~\bibnamefont
  {Hagler}},\ }\href {\doibase 10.1016/j.physrep.2009.12.008} {\bibfield
  {journal} {\bibinfo  {journal} {Phys. Rept.}\ }\textbf {\bibinfo {volume}
  {490}},\ \bibinfo {pages} {49} (\bibinfo {year} {2010})},\ \Eprint
  {http://arxiv.org/abs/0912.5483} {arXiv:0912.5483 [hep-lat]} \BibitemShut
  {NoStop}%
\bibitem [{\citenamefont {Ji}(2013)}]{Ji:2013dva}%
  \BibitemOpen
  \bibfield  {author} {\bibinfo {author} {\bibfnamefont {X.}~\bibnamefont
  {Ji}},\ }\href {\doibase 10.1103/PhysRevLett.110.262002} {\bibfield
  {journal} {\bibinfo  {journal} {Phys. Rev. Lett.}\ }\textbf {\bibinfo
  {volume} {110}},\ \bibinfo {pages} {262002} (\bibinfo {year} {2013})},\
  \Eprint {http://arxiv.org/abs/1305.1539} {arXiv:1305.1539 [hep-ph]}
  \BibitemShut {NoStop}%
\bibitem [{\citenamefont {Ji}(2014)}]{Ji:2014gla}%
  \BibitemOpen
  \bibfield  {author} {\bibinfo {author} {\bibfnamefont {X.}~\bibnamefont
  {Ji}},\ }\href {\doibase 10.1007/s11433-014-5492-3} {\bibfield  {journal}
  {\bibinfo  {journal} {Sci. China Phys. Mech. Astron.}\ }\textbf {\bibinfo
  {volume} {57}},\ \bibinfo {pages} {1407} (\bibinfo {year} {2014})},\ \Eprint
  {http://arxiv.org/abs/1404.6680} {arXiv:1404.6680 [hep-ph]} \BibitemShut
  {NoStop}%
\bibitem [{\citenamefont {Ma}\ and\ \citenamefont {Qiu}(2018)}]{Ma:2017pxb}%
  \BibitemOpen
  \bibfield  {author} {\bibinfo {author} {\bibfnamefont {Y.-Q.}\ \bibnamefont
  {Ma}}\ and\ \bibinfo {author} {\bibfnamefont {J.-W.}\ \bibnamefont {Qiu}},\
  }\href {\doibase 10.1103/PhysRevLett.120.022003} {\bibfield  {journal}
  {\bibinfo  {journal} {Phys. Rev. Lett.}\ }\textbf {\bibinfo {volume} {120}},\
  \bibinfo {pages} {022003} (\bibinfo {year} {2018})},\ \Eprint
  {http://arxiv.org/abs/1709.03018} {arXiv:1709.03018 [hep-ph]} \BibitemShut
  {NoStop}%
\bibitem [{\citenamefont {Izubuchi}\ \emph {et~al.}(2018)\citenamefont
  {Izubuchi}, \citenamefont {Ji}, \citenamefont {Jin}, \citenamefont
  {Stewart},\ and\ \citenamefont {Zhao}}]{Izubuchi:2018srq}%
  \BibitemOpen
  \bibfield  {author} {\bibinfo {author} {\bibfnamefont {T.}~\bibnamefont
  {Izubuchi}}, \bibinfo {author} {\bibfnamefont {X.}~\bibnamefont {Ji}},
  \bibinfo {author} {\bibfnamefont {L.}~\bibnamefont {Jin}}, \bibinfo {author}
  {\bibfnamefont {I.~W.}\ \bibnamefont {Stewart}}, \ and\ \bibinfo {author}
  {\bibfnamefont {Y.}~\bibnamefont {Zhao}},\ }\href {\doibase
  10.1103/PhysRevD.98.056004} {\bibfield  {journal} {\bibinfo  {journal} {Phys.
  Rev.}\ }\textbf {\bibinfo {volume} {D98}},\ \bibinfo {pages} {056004}
  (\bibinfo {year} {2018})},\ \Eprint {http://arxiv.org/abs/1801.03917}
  {arXiv:1801.03917 [hep-ph]} \BibitemShut {NoStop}%
\bibitem [{\citenamefont {Liu}\ \emph {et~al.}(2019)\citenamefont {Liu},
  \citenamefont {Wang}, \citenamefont {Xu}, \citenamefont {Zhang},
  \citenamefont {Zhang}, \citenamefont {Zhao},\ and\ \citenamefont
  {Zhao}}]{Liu:2019urm}%
  \BibitemOpen
  \bibfield  {author} {\bibinfo {author} {\bibfnamefont {Y.-S.}\ \bibnamefont
  {Liu}}, \bibinfo {author} {\bibfnamefont {W.}~\bibnamefont {Wang}}, \bibinfo
  {author} {\bibfnamefont {J.}~\bibnamefont {Xu}}, \bibinfo {author}
  {\bibfnamefont {Q.-A.}\ \bibnamefont {Zhang}}, \bibinfo {author}
  {\bibfnamefont {J.-H.}\ \bibnamefont {Zhang}}, \bibinfo {author}
  {\bibfnamefont {S.}~\bibnamefont {Zhao}}, \ and\ \bibinfo {author}
  {\bibfnamefont {Y.}~\bibnamefont {Zhao}},\ }\href@noop {} {\  (\bibinfo
  {year} {2019})},\ \Eprint {http://arxiv.org/abs/1902.00307} {arXiv:1902.00307
  [hep-ph]} \BibitemShut {NoStop}%
\bibitem [{\citenamefont {Xiong}\ \emph {et~al.}(2014)\citenamefont {Xiong},
  \citenamefont {Ji}, \citenamefont {Zhang},\ and\ \citenamefont
  {Zhao}}]{Xiong:2013bka}%
  \BibitemOpen
  \bibfield  {author} {\bibinfo {author} {\bibfnamefont {X.}~\bibnamefont
  {Xiong}}, \bibinfo {author} {\bibfnamefont {X.}~\bibnamefont {Ji}}, \bibinfo
  {author} {\bibfnamefont {J.-H.}\ \bibnamefont {Zhang}}, \ and\ \bibinfo
  {author} {\bibfnamefont {Y.}~\bibnamefont {Zhao}},\ }\href {\doibase
  10.1103/PhysRevD.90.014051} {\bibfield  {journal} {\bibinfo  {journal} {Phys.
  Rev.}\ }\textbf {\bibinfo {volume} {D90}},\ \bibinfo {pages} {014051}
  (\bibinfo {year} {2014})},\ \Eprint {http://arxiv.org/abs/1310.7471}
  {arXiv:1310.7471 [hep-ph]} \BibitemShut {NoStop}%
\bibitem [{\citenamefont {Ji}\ and\ \citenamefont {Zhang}(2015)}]{Ji:2015jwa}%
  \BibitemOpen
  \bibfield  {author} {\bibinfo {author} {\bibfnamefont {X.}~\bibnamefont
  {Ji}}\ and\ \bibinfo {author} {\bibfnamefont {J.-H.}\ \bibnamefont {Zhang}},\
  }\href {\doibase 10.1103/PhysRevD.92.034006} {\bibfield  {journal} {\bibinfo
  {journal} {Phys. Rev.}\ }\textbf {\bibinfo {volume} {D92}},\ \bibinfo {pages}
  {034006} (\bibinfo {year} {2015})},\ \Eprint
  {http://arxiv.org/abs/1505.07699} {arXiv:1505.07699 [hep-ph]} \BibitemShut
  {NoStop}%
\bibitem [{\citenamefont {Ji}\ \emph {et~al.}(2015{\natexlab{a}})\citenamefont
  {Ji}, \citenamefont {Schäfer}, \citenamefont {Xiong},\ and\ \citenamefont
  {Zhang}}]{Ji:2015qla}%
  \BibitemOpen
  \bibfield  {author} {\bibinfo {author} {\bibfnamefont {X.}~\bibnamefont
  {Ji}}, \bibinfo {author} {\bibfnamefont {A.}~\bibnamefont {Schäfer}},
  \bibinfo {author} {\bibfnamefont {X.}~\bibnamefont {Xiong}}, \ and\ \bibinfo
  {author} {\bibfnamefont {J.-H.}\ \bibnamefont {Zhang}},\ }\href {\doibase
  10.1103/PhysRevD.92.014039} {\bibfield  {journal} {\bibinfo  {journal} {Phys.
  Rev.}\ }\textbf {\bibinfo {volume} {D92}},\ \bibinfo {pages} {014039}
  (\bibinfo {year} {2015}{\natexlab{a}})},\ \Eprint
  {http://arxiv.org/abs/1506.00248} {arXiv:1506.00248 [hep-ph]} \BibitemShut
  {NoStop}%
\bibitem [{\citenamefont {Xiong}\ and\ \citenamefont
  {Zhang}(2015)}]{Xiong:2015nua}%
  \BibitemOpen
  \bibfield  {author} {\bibinfo {author} {\bibfnamefont {X.}~\bibnamefont
  {Xiong}}\ and\ \bibinfo {author} {\bibfnamefont {J.-H.}\ \bibnamefont
  {Zhang}},\ }\href {\doibase 10.1103/PhysRevD.92.054037} {\bibfield  {journal}
  {\bibinfo  {journal} {Phys. Rev.}\ }\textbf {\bibinfo {volume} {D92}},\
  \bibinfo {pages} {054037} (\bibinfo {year} {2015})},\ \Eprint
  {http://arxiv.org/abs/1509.08016} {arXiv:1509.08016 [hep-ph]} \BibitemShut
  {NoStop}%
\bibitem [{\citenamefont {Ji}\ \emph {et~al.}(2015{\natexlab{b}})\citenamefont
  {Ji}, \citenamefont {Sun}, \citenamefont {Xiong},\ and\ \citenamefont
  {Yuan}}]{Ji:2014hxa}%
  \BibitemOpen
  \bibfield  {author} {\bibinfo {author} {\bibfnamefont {X.}~\bibnamefont
  {Ji}}, \bibinfo {author} {\bibfnamefont {P.}~\bibnamefont {Sun}}, \bibinfo
  {author} {\bibfnamefont {X.}~\bibnamefont {Xiong}}, \ and\ \bibinfo {author}
  {\bibfnamefont {F.}~\bibnamefont {Yuan}},\ }\href {\doibase
  10.1103/PhysRevD.91.074009} {\bibfield  {journal} {\bibinfo  {journal} {Phys.
  Rev.}\ }\textbf {\bibinfo {volume} {D91}},\ \bibinfo {pages} {074009}
  (\bibinfo {year} {2015}{\natexlab{b}})},\ \Eprint
  {http://arxiv.org/abs/1405.7640} {arXiv:1405.7640 [hep-ph]} \BibitemShut
  {NoStop}%
\bibitem [{\citenamefont {Monahan}(2018)}]{Monahan:2017hpu}%
  \BibitemOpen
  \bibfield  {author} {\bibinfo {author} {\bibfnamefont {C.}~\bibnamefont
  {Monahan}},\ }\href {\doibase 10.1103/PhysRevD.97.054507} {\bibfield
  {journal} {\bibinfo  {journal} {Phys. Rev.}\ }\textbf {\bibinfo {volume}
  {D97}},\ \bibinfo {pages} {054507} (\bibinfo {year} {2018})},\ \Eprint
  {http://arxiv.org/abs/1710.04607} {arXiv:1710.04607 [hep-lat]} \BibitemShut
  {NoStop}%
\bibitem [{\citenamefont {Ji}\ \emph {et~al.}(2018{\natexlab{a}})\citenamefont
  {Ji}, \citenamefont {Jin}, \citenamefont {Yuan}, \citenamefont {Zhang},\ and\
  \citenamefont {Zhao}}]{Ji:2018hvs}%
  \BibitemOpen
  \bibfield  {author} {\bibinfo {author} {\bibfnamefont {X.}~\bibnamefont
  {Ji}}, \bibinfo {author} {\bibfnamefont {L.-C.}\ \bibnamefont {Jin}},
  \bibinfo {author} {\bibfnamefont {F.}~\bibnamefont {Yuan}}, \bibinfo {author}
  {\bibfnamefont {J.-H.}\ \bibnamefont {Zhang}}, \ and\ \bibinfo {author}
  {\bibfnamefont {Y.}~\bibnamefont {Zhao}},\ }\href@noop {} {\  (\bibinfo
  {year} {2018}{\natexlab{a}})},\ \Eprint {http://arxiv.org/abs/1801.05930}
  {arXiv:1801.05930 [hep-ph]} \BibitemShut {NoStop}%
\bibitem [{\citenamefont {Stewart}\ and\ \citenamefont
  {Zhao}(2018)}]{Stewart:2017tvs}%
  \BibitemOpen
  \bibfield  {author} {\bibinfo {author} {\bibfnamefont {I.~W.}\ \bibnamefont
  {Stewart}}\ and\ \bibinfo {author} {\bibfnamefont {Y.}~\bibnamefont {Zhao}},\
  }\href {\doibase 10.1103/PhysRevD.97.054512} {\bibfield  {journal} {\bibinfo
  {journal} {Phys. Rev.}\ }\textbf {\bibinfo {volume} {D97}},\ \bibinfo {pages}
  {054512} (\bibinfo {year} {2018})},\ \Eprint
  {http://arxiv.org/abs/1709.04933} {arXiv:1709.04933 [hep-ph]} \BibitemShut
  {NoStop}%
\bibitem [{\citenamefont {Constantinou}\ and\ \citenamefont
  {Panagopoulos}(2017)}]{Constantinou:2017sej}%
  \BibitemOpen
  \bibfield  {author} {\bibinfo {author} {\bibfnamefont {M.}~\bibnamefont
  {Constantinou}}\ and\ \bibinfo {author} {\bibfnamefont {H.}~\bibnamefont
  {Panagopoulos}},\ }\href {\doibase 10.1103/PhysRevD.96.054506} {\bibfield
  {journal} {\bibinfo  {journal} {Phys. Rev.}\ }\textbf {\bibinfo {volume}
  {D96}},\ \bibinfo {pages} {054506} (\bibinfo {year} {2017})},\ \Eprint
  {http://arxiv.org/abs/1705.11193} {arXiv:1705.11193 [hep-lat]} \BibitemShut
  {NoStop}%
\bibitem [{\citenamefont {Green}\ \emph {et~al.}(2018)\citenamefont {Green},
  \citenamefont {Jansen},\ and\ \citenamefont {Steffens}}]{Green:2017xeu}%
  \BibitemOpen
  \bibfield  {author} {\bibinfo {author} {\bibfnamefont {J.}~\bibnamefont
  {Green}}, \bibinfo {author} {\bibfnamefont {K.}~\bibnamefont {Jansen}}, \
  and\ \bibinfo {author} {\bibfnamefont {F.}~\bibnamefont {Steffens}},\ }\href
  {\doibase 10.1103/PhysRevLett.121.022004} {\bibfield  {journal} {\bibinfo
  {journal} {Phys. Rev. Lett.}\ }\textbf {\bibinfo {volume} {121}},\ \bibinfo
  {pages} {022004} (\bibinfo {year} {2018})},\ \Eprint
  {http://arxiv.org/abs/1707.07152} {arXiv:1707.07152 [hep-lat]} \BibitemShut
  {NoStop}%
\bibitem [{\citenamefont {Xiong}\ \emph {et~al.}(2017)\citenamefont {Xiong},
  \citenamefont {Luu},\ and\ \citenamefont {Meißner}}]{Xiong:2017jtn}%
  \BibitemOpen
  \bibfield  {author} {\bibinfo {author} {\bibfnamefont {X.}~\bibnamefont
  {Xiong}}, \bibinfo {author} {\bibfnamefont {T.}~\bibnamefont {Luu}}, \ and\
  \bibinfo {author} {\bibfnamefont {U.-G.}\ \bibnamefont {Meißner}},\
  }\href@noop {} {\  (\bibinfo {year} {2017})},\ \Eprint
  {http://arxiv.org/abs/1705.00246} {arXiv:1705.00246 [hep-ph]} \BibitemShut
  {NoStop}%
\bibitem [{\citenamefont {Wang}\ \emph {et~al.}(2018)\citenamefont {Wang},
  \citenamefont {Zhao},\ and\ \citenamefont {Zhu}}]{Wang:2017qyg}%
  \BibitemOpen
  \bibfield  {author} {\bibinfo {author} {\bibfnamefont {W.}~\bibnamefont
  {Wang}}, \bibinfo {author} {\bibfnamefont {S.}~\bibnamefont {Zhao}}, \ and\
  \bibinfo {author} {\bibfnamefont {R.}~\bibnamefont {Zhu}},\ }\href {\doibase
  10.1140/epjc/s10052-018-5617-3} {\bibfield  {journal} {\bibinfo  {journal}
  {Eur. Phys. J.}\ }\textbf {\bibinfo {volume} {C78}},\ \bibinfo {pages} {147}
  (\bibinfo {year} {2018})},\ \Eprint {http://arxiv.org/abs/1708.02458}
  {arXiv:1708.02458 [hep-ph]} \BibitemShut {NoStop}%
\bibitem [{\citenamefont {Wang}\ and\ \citenamefont
  {Zhao}(2018)}]{Wang:2017eel}%
  \BibitemOpen
  \bibfield  {author} {\bibinfo {author} {\bibfnamefont {W.}~\bibnamefont
  {Wang}}\ and\ \bibinfo {author} {\bibfnamefont {S.}~\bibnamefont {Zhao}},\
  }\href {\doibase 10.1007/JHEP05(2018)142} {\bibfield  {journal} {\bibinfo
  {journal} {JHEP}\ }\textbf {\bibinfo {volume} {05}},\ \bibinfo {pages} {142}
  (\bibinfo {year} {2018})},\ \Eprint {http://arxiv.org/abs/1712.09247}
  {arXiv:1712.09247 [hep-ph]} \BibitemShut {NoStop}%
\bibitem [{\citenamefont {Xu}\ \emph {et~al.}(2018{\natexlab{a}})\citenamefont
  {Xu}, \citenamefont {Zhang},\ and\ \citenamefont {Zhao}}]{Xu:2018mpf}%
  \BibitemOpen
  \bibfield  {author} {\bibinfo {author} {\bibfnamefont {J.}~\bibnamefont
  {Xu}}, \bibinfo {author} {\bibfnamefont {Q.-A.}\ \bibnamefont {Zhang}}, \
  and\ \bibinfo {author} {\bibfnamefont {S.}~\bibnamefont {Zhao}},\ }\href
  {\doibase 10.1103/PhysRevD.97.114026} {\bibfield  {journal} {\bibinfo
  {journal} {Phys. Rev.}\ }\textbf {\bibinfo {volume} {D97}},\ \bibinfo {pages}
  {114026} (\bibinfo {year} {2018}{\natexlab{a}})},\ \Eprint
  {http://arxiv.org/abs/1804.01042} {arXiv:1804.01042 [hep-ph]} \BibitemShut
  {NoStop}%
\bibitem [{\citenamefont {Chen}\ \emph {et~al.}(2016)\citenamefont {Chen},
  \citenamefont {Cohen}, \citenamefont {Ji}, \citenamefont {Lin},\ and\
  \citenamefont {Zhang}}]{Chen:2016utp}%
  \BibitemOpen
  \bibfield  {author} {\bibinfo {author} {\bibfnamefont {J.-W.}\ \bibnamefont
  {Chen}}, \bibinfo {author} {\bibfnamefont {S.~D.}\ \bibnamefont {Cohen}},
  \bibinfo {author} {\bibfnamefont {X.}~\bibnamefont {Ji}}, \bibinfo {author}
  {\bibfnamefont {H.-W.}\ \bibnamefont {Lin}}, \ and\ \bibinfo {author}
  {\bibfnamefont {J.-H.}\ \bibnamefont {Zhang}},\ }\href {\doibase
  10.1016/j.nuclphysb.2016.07.033} {\bibfield  {journal} {\bibinfo  {journal}
  {Nucl. Phys.}\ }\textbf {\bibinfo {volume} {B911}},\ \bibinfo {pages} {246}
  (\bibinfo {year} {2016})},\ \Eprint {http://arxiv.org/abs/1603.06664}
  {arXiv:1603.06664 [hep-ph]} \BibitemShut {NoStop}%
\bibitem [{\citenamefont {Zhang}\ \emph {et~al.}(2017)\citenamefont {Zhang},
  \citenamefont {Chen}, \citenamefont {Ji}, \citenamefont {Jin},\ and\
  \citenamefont {Lin}}]{Zhang:2017bzy}%
  \BibitemOpen
  \bibfield  {author} {\bibinfo {author} {\bibfnamefont {J.-H.}\ \bibnamefont
  {Zhang}}, \bibinfo {author} {\bibfnamefont {J.-W.}\ \bibnamefont {Chen}},
  \bibinfo {author} {\bibfnamefont {X.}~\bibnamefont {Ji}}, \bibinfo {author}
  {\bibfnamefont {L.}~\bibnamefont {Jin}}, \ and\ \bibinfo {author}
  {\bibfnamefont {H.-W.}\ \bibnamefont {Lin}},\ }\href {\doibase
  10.1103/PhysRevD.95.094514} {\bibfield  {journal} {\bibinfo  {journal} {Phys.
  Rev.}\ }\textbf {\bibinfo {volume} {D95}},\ \bibinfo {pages} {094514}
  (\bibinfo {year} {2017})},\ \Eprint {http://arxiv.org/abs/1702.00008}
  {arXiv:1702.00008 [hep-lat]} \BibitemShut {NoStop}%
\bibitem [{\citenamefont {Ishikawa}\ \emph {et~al.}(2016)\citenamefont
  {Ishikawa}, \citenamefont {Ma}, \citenamefont {Qiu},\ and\ \citenamefont
  {Yoshida}}]{Ishikawa:2016znu}%
  \BibitemOpen
  \bibfield  {author} {\bibinfo {author} {\bibfnamefont {T.}~\bibnamefont
  {Ishikawa}}, \bibinfo {author} {\bibfnamefont {Y.-Q.}\ \bibnamefont {Ma}},
  \bibinfo {author} {\bibfnamefont {J.-W.}\ \bibnamefont {Qiu}}, \ and\
  \bibinfo {author} {\bibfnamefont {S.}~\bibnamefont {Yoshida}},\ }\href@noop
  {} {\  (\bibinfo {year} {2016})},\ \Eprint {http://arxiv.org/abs/1609.02018}
  {arXiv:1609.02018 [hep-lat]} \BibitemShut {NoStop}%
\bibitem [{\citenamefont {Chen}\ \emph
  {et~al.}(2017{\natexlab{a}})\citenamefont {Chen}, \citenamefont {Ji},\ and\
  \citenamefont {Zhang}}]{Chen:2016fxx}%
  \BibitemOpen
  \bibfield  {author} {\bibinfo {author} {\bibfnamefont {J.-W.}\ \bibnamefont
  {Chen}}, \bibinfo {author} {\bibfnamefont {X.}~\bibnamefont {Ji}}, \ and\
  \bibinfo {author} {\bibfnamefont {J.-H.}\ \bibnamefont {Zhang}},\ }\href
  {\doibase 10.1016/j.nuclphysb.2016.12.004} {\bibfield  {journal} {\bibinfo
  {journal} {Nucl. Phys.}\ }\textbf {\bibinfo {volume} {B915}},\ \bibinfo
  {pages} {1} (\bibinfo {year} {2017}{\natexlab{a}})},\ \Eprint
  {http://arxiv.org/abs/1609.08102} {arXiv:1609.08102 [hep-ph]} \BibitemShut
  {NoStop}%
\bibitem [{\citenamefont {Ji}\ \emph {et~al.}(2018{\natexlab{b}})\citenamefont
  {Ji}, \citenamefont {Zhang},\ and\ \citenamefont {Zhao}}]{Ji:2017oey}%
  \BibitemOpen
  \bibfield  {author} {\bibinfo {author} {\bibfnamefont {X.}~\bibnamefont
  {Ji}}, \bibinfo {author} {\bibfnamefont {J.-H.}\ \bibnamefont {Zhang}}, \
  and\ \bibinfo {author} {\bibfnamefont {Y.}~\bibnamefont {Zhao}},\ }\href
  {\doibase 10.1103/PhysRevLett.120.112001} {\bibfield  {journal} {\bibinfo
  {journal} {Phys. Rev. Lett.}\ }\textbf {\bibinfo {volume} {120}},\ \bibinfo
  {pages} {112001} (\bibinfo {year} {2018}{\natexlab{b}})},\ \Eprint
  {http://arxiv.org/abs/1706.08962} {arXiv:1706.08962 [hep-ph]} \BibitemShut
  {NoStop}%
\bibitem [{\citenamefont {Ishikawa}\ \emph {et~al.}(2017)\citenamefont
  {Ishikawa}, \citenamefont {Ma}, \citenamefont {Qiu},\ and\ \citenamefont
  {Yoshida}}]{Ishikawa:2017faj}%
  \BibitemOpen
  \bibfield  {author} {\bibinfo {author} {\bibfnamefont {T.}~\bibnamefont
  {Ishikawa}}, \bibinfo {author} {\bibfnamefont {Y.-Q.}\ \bibnamefont {Ma}},
  \bibinfo {author} {\bibfnamefont {J.-W.}\ \bibnamefont {Qiu}}, \ and\
  \bibinfo {author} {\bibfnamefont {S.}~\bibnamefont {Yoshida}},\ }\href
  {\doibase 10.1103/PhysRevD.96.094019} {\bibfield  {journal} {\bibinfo
  {journal} {Phys. Rev.}\ }\textbf {\bibinfo {volume} {D96}},\ \bibinfo {pages}
  {094019} (\bibinfo {year} {2017})},\ \Eprint
  {http://arxiv.org/abs/1707.03107} {arXiv:1707.03107 [hep-ph]} \BibitemShut
  {NoStop}%
\bibitem [{\citenamefont {Chen}\ \emph
  {et~al.}(2018{\natexlab{a}})\citenamefont {Chen}, \citenamefont {Ishikawa},
  \citenamefont {Jin}, \citenamefont {Lin}, \citenamefont {Yang}, \citenamefont
  {Zhang},\ and\ \citenamefont {Zhao}}]{Chen:2017mzz}%
  \BibitemOpen
  \bibfield  {author} {\bibinfo {author} {\bibfnamefont {J.-W.}\ \bibnamefont
  {Chen}}, \bibinfo {author} {\bibfnamefont {T.}~\bibnamefont {Ishikawa}},
  \bibinfo {author} {\bibfnamefont {L.}~\bibnamefont {Jin}}, \bibinfo {author}
  {\bibfnamefont {H.-W.}\ \bibnamefont {Lin}}, \bibinfo {author} {\bibfnamefont
  {Y.-B.}\ \bibnamefont {Yang}}, \bibinfo {author} {\bibfnamefont {J.-H.}\
  \bibnamefont {Zhang}}, \ and\ \bibinfo {author} {\bibfnamefont
  {Y.}~\bibnamefont {Zhao}},\ }\href {\doibase 10.1103/PhysRevD.97.014505}
  {\bibfield  {journal} {\bibinfo  {journal} {Phys. Rev.}\ }\textbf {\bibinfo
  {volume} {D97}},\ \bibinfo {pages} {014505} (\bibinfo {year}
  {2018}{\natexlab{a}})},\ \Eprint {http://arxiv.org/abs/1706.01295}
  {arXiv:1706.01295 [hep-lat]} \BibitemShut {NoStop}%
\bibitem [{\citenamefont {Alexandrou}\ \emph
  {et~al.}(2017{\natexlab{a}})\citenamefont {Alexandrou}, \citenamefont
  {Cichy}, \citenamefont {Constantinou}, \citenamefont {Hadjiyiannakou},
  \citenamefont {Jansen}, \citenamefont {Panagopoulos},\ and\ \citenamefont
  {Steffens}}]{Alexandrou:2017huk}%
  \BibitemOpen
  \bibfield  {author} {\bibinfo {author} {\bibfnamefont {C.}~\bibnamefont
  {Alexandrou}}, \bibinfo {author} {\bibfnamefont {K.}~\bibnamefont {Cichy}},
  \bibinfo {author} {\bibfnamefont {M.}~\bibnamefont {Constantinou}}, \bibinfo
  {author} {\bibfnamefont {K.}~\bibnamefont {Hadjiyiannakou}}, \bibinfo
  {author} {\bibfnamefont {K.}~\bibnamefont {Jansen}}, \bibinfo {author}
  {\bibfnamefont {H.}~\bibnamefont {Panagopoulos}}, \ and\ \bibinfo {author}
  {\bibfnamefont {F.}~\bibnamefont {Steffens}},\ }\href {\doibase
  10.1016/j.nuclphysb.2017.08.012} {\bibfield  {journal} {\bibinfo  {journal}
  {Nucl. Phys.}\ }\textbf {\bibinfo {volume} {B923}},\ \bibinfo {pages} {394}
  (\bibinfo {year} {2017}{\natexlab{a}})},\ \Eprint
  {http://arxiv.org/abs/1706.00265} {arXiv:1706.00265 [hep-lat]} \BibitemShut
  {NoStop}%
\bibitem [{\citenamefont {Chen}\ \emph
  {et~al.}(2017{\natexlab{b}})\citenamefont {Chen}, \citenamefont {Ishikawa},
  \citenamefont {Jin}, \citenamefont {Lin}, \citenamefont {Yang}, \citenamefont
  {Zhang},\ and\ \citenamefont {Zhao}}]{Chen:2017mie}%
  \BibitemOpen
  \bibfield  {author} {\bibinfo {author} {\bibfnamefont {J.-W.}\ \bibnamefont
  {Chen}}, \bibinfo {author} {\bibfnamefont {T.}~\bibnamefont {Ishikawa}},
  \bibinfo {author} {\bibfnamefont {L.}~\bibnamefont {Jin}}, \bibinfo {author}
  {\bibfnamefont {H.-W.}\ \bibnamefont {Lin}}, \bibinfo {author} {\bibfnamefont
  {Y.-B.}\ \bibnamefont {Yang}}, \bibinfo {author} {\bibfnamefont {J.-H.}\
  \bibnamefont {Zhang}}, \ and\ \bibinfo {author} {\bibfnamefont
  {Y.}~\bibnamefont {Zhao}},\ }\href@noop {} {\  (\bibinfo {year}
  {2017}{\natexlab{b}})},\ \Eprint {http://arxiv.org/abs/1710.01089}
  {arXiv:1710.01089 [hep-lat]} \BibitemShut {NoStop}%
\bibitem [{\citenamefont {Lin}\ \emph {et~al.}(2018{\natexlab{a}})\citenamefont
  {Lin}, \citenamefont {Chen}, \citenamefont {Ishikawa},\ and\ \citenamefont
  {Zhang}}]{Lin:2017ani}%
  \BibitemOpen
  \bibfield  {author} {\bibinfo {author} {\bibfnamefont {H.-W.}\ \bibnamefont
  {Lin}}, \bibinfo {author} {\bibfnamefont {J.-W.}\ \bibnamefont {Chen}},
  \bibinfo {author} {\bibfnamefont {T.}~\bibnamefont {Ishikawa}}, \ and\
  \bibinfo {author} {\bibfnamefont {J.-H.}\ \bibnamefont {Zhang}} (\bibinfo
  {collaboration} {LP3}),\ }\href {\doibase 10.1103/PhysRevD.98.054504}
  {\bibfield  {journal} {\bibinfo  {journal} {Phys. Rev.}\ }\textbf {\bibinfo
  {volume} {D98}},\ \bibinfo {pages} {054504} (\bibinfo {year}
  {2018}{\natexlab{a}})},\ \Eprint {http://arxiv.org/abs/1708.05301}
  {arXiv:1708.05301 [hep-lat]} \BibitemShut {NoStop}%
\bibitem [{\citenamefont {Chen}\ \emph
  {et~al.}(2017{\natexlab{c}})\citenamefont {Chen}, \citenamefont {Ishikawa},
  \citenamefont {Jin}, \citenamefont {Lin}, \citenamefont {Schäfer},
  \citenamefont {Yang}, \citenamefont {Zhang},\ and\ \citenamefont
  {Zhao}}]{Chen:2017lnm}%
  \BibitemOpen
  \bibfield  {author} {\bibinfo {author} {\bibfnamefont {J.-W.}\ \bibnamefont
  {Chen}}, \bibinfo {author} {\bibfnamefont {T.}~\bibnamefont {Ishikawa}},
  \bibinfo {author} {\bibfnamefont {L.}~\bibnamefont {Jin}}, \bibinfo {author}
  {\bibfnamefont {H.-W.}\ \bibnamefont {Lin}}, \bibinfo {author} {\bibfnamefont
  {A.}~\bibnamefont {Schäfer}}, \bibinfo {author} {\bibfnamefont {Y.-B.}\
  \bibnamefont {Yang}}, \bibinfo {author} {\bibfnamefont {J.-H.}\ \bibnamefont
  {Zhang}}, \ and\ \bibinfo {author} {\bibfnamefont {Y.}~\bibnamefont {Zhao}},\
  }\href@noop {} {\  (\bibinfo {year} {2017}{\natexlab{c}})},\ \Eprint
  {http://arxiv.org/abs/1711.07858} {arXiv:1711.07858 [hep-ph]} \BibitemShut
  {NoStop}%
\bibitem [{\citenamefont {Li}(2016)}]{Li:2016amo}%
  \BibitemOpen
  \bibfield  {author} {\bibinfo {author} {\bibfnamefont {H.-n.}\ \bibnamefont
  {Li}},\ }\href {\doibase 10.1103/PhysRevD.94.074036} {\bibfield  {journal}
  {\bibinfo  {journal} {Phys. Rev.}\ }\textbf {\bibinfo {volume} {D94}},\
  \bibinfo {pages} {074036} (\bibinfo {year} {2016})},\ \Eprint
  {http://arxiv.org/abs/1602.07575} {arXiv:1602.07575 [hep-ph]} \BibitemShut
  {NoStop}%
\bibitem [{\citenamefont {Monahan}\ and\ \citenamefont
  {Orginos}(2017)}]{Monahan:2016bvm}%
  \BibitemOpen
  \bibfield  {author} {\bibinfo {author} {\bibfnamefont {C.}~\bibnamefont
  {Monahan}}\ and\ \bibinfo {author} {\bibfnamefont {K.}~\bibnamefont
  {Orginos}},\ }\href {\doibase 10.1007/JHEP03(2017)116} {\bibfield  {journal}
  {\bibinfo  {journal} {JHEP}\ }\textbf {\bibinfo {volume} {03}},\ \bibinfo
  {pages} {116} (\bibinfo {year} {2017})},\ \Eprint
  {http://arxiv.org/abs/1612.01584} {arXiv:1612.01584 [hep-lat]} \BibitemShut
  {NoStop}%
\bibitem [{\citenamefont {Radyushkin}(2017)}]{Radyushkin:2016hsy}%
  \BibitemOpen
  \bibfield  {author} {\bibinfo {author} {\bibfnamefont {A.}~\bibnamefont
  {Radyushkin}},\ }\href {\doibase 10.1016/j.physletb.2017.02.019} {\bibfield
  {journal} {\bibinfo  {journal} {Phys. Lett.}\ }\textbf {\bibinfo {volume}
  {B767}},\ \bibinfo {pages} {314} (\bibinfo {year} {2017})},\ \Eprint
  {http://arxiv.org/abs/1612.05170} {arXiv:1612.05170 [hep-ph]} \BibitemShut
  {NoStop}%
\bibitem [{\citenamefont {Rossi}\ and\ \citenamefont
  {Testa}(2017)}]{Rossi:2017muf}%
  \BibitemOpen
  \bibfield  {author} {\bibinfo {author} {\bibfnamefont {G.~C.}\ \bibnamefont
  {Rossi}}\ and\ \bibinfo {author} {\bibfnamefont {M.}~\bibnamefont {Testa}},\
  }\href {\doibase 10.1103/PhysRevD.96.014507} {\bibfield  {journal} {\bibinfo
  {journal} {Phys. Rev.}\ }\textbf {\bibinfo {volume} {D96}},\ \bibinfo {pages}
  {014507} (\bibinfo {year} {2017})},\ \Eprint
  {http://arxiv.org/abs/1706.04428} {arXiv:1706.04428 [hep-lat]} \BibitemShut
  {NoStop}%
\bibitem [{\citenamefont {Carlson}\ and\ \citenamefont
  {Freid}(2017)}]{Carlson:2017gpk}%
  \BibitemOpen
  \bibfield  {author} {\bibinfo {author} {\bibfnamefont {C.~E.}\ \bibnamefont
  {Carlson}}\ and\ \bibinfo {author} {\bibfnamefont {M.}~\bibnamefont
  {Freid}},\ }\href {\doibase 10.1103/PhysRevD.95.094504} {\bibfield  {journal}
  {\bibinfo  {journal} {Phys. Rev.}\ }\textbf {\bibinfo {volume} {D95}},\
  \bibinfo {pages} {094504} (\bibinfo {year} {2017})},\ \Eprint
  {http://arxiv.org/abs/1702.05775} {arXiv:1702.05775 [hep-ph]} \BibitemShut
  {NoStop}%
\bibitem [{\citenamefont {Ji}\ \emph {et~al.}(2017)\citenamefont {Ji},
  \citenamefont {Zhang},\ and\ \citenamefont {Zhao}}]{Ji:2017rah}%
  \BibitemOpen
  \bibfield  {author} {\bibinfo {author} {\bibfnamefont {X.}~\bibnamefont
  {Ji}}, \bibinfo {author} {\bibfnamefont {J.-H.}\ \bibnamefont {Zhang}}, \
  and\ \bibinfo {author} {\bibfnamefont {Y.}~\bibnamefont {Zhao}},\ }\href
  {\doibase 10.1016/j.nuclphysb.2017.09.001} {\bibfield  {journal} {\bibinfo
  {journal} {Nucl. Phys.}\ }\textbf {\bibinfo {volume} {B924}},\ \bibinfo
  {pages} {366} (\bibinfo {year} {2017})},\ \Eprint
  {http://arxiv.org/abs/1706.07416} {arXiv:1706.07416 [hep-ph]} \BibitemShut
  {NoStop}%
\bibitem [{\citenamefont {Briceño}\ \emph {et~al.}(2018)\citenamefont
  {Briceño}, \citenamefont {Guerrero}, \citenamefont {Hansen},\ and\
  \citenamefont {Monahan}}]{Briceno:2018lfj}%
  \BibitemOpen
  \bibfield  {author} {\bibinfo {author} {\bibfnamefont {R.~A.}\ \bibnamefont
  {Briceño}}, \bibinfo {author} {\bibfnamefont {J.~V.}\ \bibnamefont
  {Guerrero}}, \bibinfo {author} {\bibfnamefont {M.~T.}\ \bibnamefont
  {Hansen}}, \ and\ \bibinfo {author} {\bibfnamefont {C.~J.}\ \bibnamefont
  {Monahan}},\ }\href {\doibase 10.1103/PhysRevD.98.014511} {\bibfield
  {journal} {\bibinfo  {journal} {Phys. Rev.}\ }\textbf {\bibinfo {volume}
  {D98}},\ \bibinfo {pages} {014511} (\bibinfo {year} {2018})},\ \Eprint
  {http://arxiv.org/abs/1805.01034} {arXiv:1805.01034 [hep-lat]} \BibitemShut
  {NoStop}%
\bibitem [{\citenamefont {Hobbs}(2018)}]{Hobbs:2017xtq}%
  \BibitemOpen
  \bibfield  {author} {\bibinfo {author} {\bibfnamefont {T.~J.}\ \bibnamefont
  {Hobbs}},\ }\href {\doibase 10.1103/PhysRevD.97.054028} {\bibfield  {journal}
  {\bibinfo  {journal} {Phys. Rev.}\ }\textbf {\bibinfo {volume} {D97}},\
  \bibinfo {pages} {054028} (\bibinfo {year} {2018})},\ \Eprint
  {http://arxiv.org/abs/1708.05463} {arXiv:1708.05463 [hep-ph]} \BibitemShut
  {NoStop}%
\bibitem [{\citenamefont {Jia}\ \emph {et~al.}(2017)\citenamefont {Jia},
  \citenamefont {Liang}, \citenamefont {Li},\ and\ \citenamefont
  {Xiong}}]{Jia:2017uul}%
  \BibitemOpen
  \bibfield  {author} {\bibinfo {author} {\bibfnamefont {Y.}~\bibnamefont
  {Jia}}, \bibinfo {author} {\bibfnamefont {S.}~\bibnamefont {Liang}}, \bibinfo
  {author} {\bibfnamefont {L.}~\bibnamefont {Li}}, \ and\ \bibinfo {author}
  {\bibfnamefont {X.}~\bibnamefont {Xiong}},\ }\href {\doibase
  10.1007/JHEP11(2017)151} {\bibfield  {journal} {\bibinfo  {journal} {JHEP}\
  }\textbf {\bibinfo {volume} {11}},\ \bibinfo {pages} {151} (\bibinfo {year}
  {2017})},\ \Eprint {http://arxiv.org/abs/1708.09379} {arXiv:1708.09379
  [hep-ph]} \BibitemShut {NoStop}%
\bibitem [{\citenamefont {Xu}\ \emph {et~al.}(2018{\natexlab{b}})\citenamefont
  {Xu}, \citenamefont {Chang}, \citenamefont {Roberts},\ and\ \citenamefont
  {Zong}}]{Xu:2018eii}%
  \BibitemOpen
  \bibfield  {author} {\bibinfo {author} {\bibfnamefont {S.-S.}\ \bibnamefont
  {Xu}}, \bibinfo {author} {\bibfnamefont {L.}~\bibnamefont {Chang}}, \bibinfo
  {author} {\bibfnamefont {C.~D.}\ \bibnamefont {Roberts}}, \ and\ \bibinfo
  {author} {\bibfnamefont {H.-S.}\ \bibnamefont {Zong}},\ }\href {\doibase
  10.1103/PhysRevD.97.094014} {\bibfield  {journal} {\bibinfo  {journal} {Phys.
  Rev.}\ }\textbf {\bibinfo {volume} {D97}},\ \bibinfo {pages} {094014}
  (\bibinfo {year} {2018}{\natexlab{b}})},\ \Eprint
  {http://arxiv.org/abs/1802.09552} {arXiv:1802.09552 [nucl-th]} \BibitemShut
  {NoStop}%
\bibitem [{\citenamefont {Jia}\ \emph {et~al.}(2018)\citenamefont {Jia},
  \citenamefont {Liang}, \citenamefont {Xiong},\ and\ \citenamefont
  {Yu}}]{Jia:2018qee}%
  \BibitemOpen
  \bibfield  {author} {\bibinfo {author} {\bibfnamefont {Y.}~\bibnamefont
  {Jia}}, \bibinfo {author} {\bibfnamefont {S.}~\bibnamefont {Liang}}, \bibinfo
  {author} {\bibfnamefont {X.}~\bibnamefont {Xiong}}, \ and\ \bibinfo {author}
  {\bibfnamefont {R.}~\bibnamefont {Yu}},\ }\href {\doibase
  10.1103/PhysRevD.98.054011} {\bibfield  {journal} {\bibinfo  {journal} {Phys.
  Rev.}\ }\textbf {\bibinfo {volume} {D98}},\ \bibinfo {pages} {054011}
  (\bibinfo {year} {2018})},\ \Eprint {http://arxiv.org/abs/1804.04644}
  {arXiv:1804.04644 [hep-th]} \BibitemShut {NoStop}%
\bibitem [{\citenamefont {Spanoudes}\ and\ \citenamefont
  {Panagopoulos}(2018)}]{Spanoudes:2018zya}%
  \BibitemOpen
  \bibfield  {author} {\bibinfo {author} {\bibfnamefont {G.}~\bibnamefont
  {Spanoudes}}\ and\ \bibinfo {author} {\bibfnamefont {H.}~\bibnamefont
  {Panagopoulos}},\ }\href {\doibase 10.1103/PhysRevD.98.014509} {\bibfield
  {journal} {\bibinfo  {journal} {Phys. Rev.}\ }\textbf {\bibinfo {volume}
  {D98}},\ \bibinfo {pages} {014509} (\bibinfo {year} {2018})},\ \Eprint
  {http://arxiv.org/abs/1805.01164} {arXiv:1805.01164 [hep-lat]} \BibitemShut
  {NoStop}%
\bibitem [{\citenamefont {Rossi}\ and\ \citenamefont
  {Testa}(2018)}]{Rossi:2018zkn}%
  \BibitemOpen
  \bibfield  {author} {\bibinfo {author} {\bibfnamefont {G.}~\bibnamefont
  {Rossi}}\ and\ \bibinfo {author} {\bibfnamefont {M.}~\bibnamefont {Testa}},\
  }\href {\doibase 10.1103/PhysRevD.98.054028} {\bibfield  {journal} {\bibinfo
  {journal} {Phys. Rev.}\ }\textbf {\bibinfo {volume} {D98}},\ \bibinfo {pages}
  {054028} (\bibinfo {year} {2018})},\ \Eprint
  {http://arxiv.org/abs/1806.00808} {arXiv:1806.00808 [hep-lat]} \BibitemShut
  {NoStop}%
\bibitem [{\citenamefont {Liu}\ \emph {et~al.}(2018{\natexlab{a}})\citenamefont
  {Liu}, \citenamefont {Chen}, \citenamefont {Jin}, \citenamefont {Lin},
  \citenamefont {Yang}, \citenamefont {Zhang},\ and\ \citenamefont
  {Zhao}}]{Liu:2018uuj}%
  \BibitemOpen
  \bibfield  {author} {\bibinfo {author} {\bibfnamefont {Y.-S.}\ \bibnamefont
  {Liu}}, \bibinfo {author} {\bibfnamefont {J.-W.}\ \bibnamefont {Chen}},
  \bibinfo {author} {\bibfnamefont {L.}~\bibnamefont {Jin}}, \bibinfo {author}
  {\bibfnamefont {H.-W.}\ \bibnamefont {Lin}}, \bibinfo {author} {\bibfnamefont
  {Y.-B.}\ \bibnamefont {Yang}}, \bibinfo {author} {\bibfnamefont {J.-H.}\
  \bibnamefont {Zhang}}, \ and\ \bibinfo {author} {\bibfnamefont
  {Y.}~\bibnamefont {Zhao}},\ }\href@noop {} {\  (\bibinfo {year}
  {2018}{\natexlab{a}})},\ \Eprint {http://arxiv.org/abs/1807.06566}
  {arXiv:1807.06566 [hep-lat]} \BibitemShut {NoStop}%
\bibitem [{\citenamefont {Ji}\ \emph {et~al.}(2019)\citenamefont {Ji},
  \citenamefont {Liu},\ and\ \citenamefont {Zahed}}]{Ji:2018waw}%
  \BibitemOpen
  \bibfield  {author} {\bibinfo {author} {\bibfnamefont {X.}~\bibnamefont
  {Ji}}, \bibinfo {author} {\bibfnamefont {Y.}~\bibnamefont {Liu}}, \ and\
  \bibinfo {author} {\bibfnamefont {I.}~\bibnamefont {Zahed}},\ }\href
  {\doibase 10.1103/PhysRevD.99.054008} {\bibfield  {journal} {\bibinfo
  {journal} {Phys. Rev.}\ }\textbf {\bibinfo {volume} {D99}},\ \bibinfo {pages}
  {054008} (\bibinfo {year} {2019})},\ \Eprint
  {http://arxiv.org/abs/1807.07528} {arXiv:1807.07528 [hep-ph]} \BibitemShut
  {NoStop}%
\bibitem [{\citenamefont {Bhattacharya}\ \emph
  {et~al.}(2019{\natexlab{a}})\citenamefont {Bhattacharya}, \citenamefont
  {Cocuzza},\ and\ \citenamefont {Metz}}]{Bhattacharya:2018zxi}%
  \BibitemOpen
  \bibfield  {author} {\bibinfo {author} {\bibfnamefont {S.}~\bibnamefont
  {Bhattacharya}}, \bibinfo {author} {\bibfnamefont {C.}~\bibnamefont
  {Cocuzza}}, \ and\ \bibinfo {author} {\bibfnamefont {A.}~\bibnamefont
  {Metz}},\ }\href {\doibase 10.1016/j.physletb.2018.09.061} {\bibfield
  {journal} {\bibinfo  {journal} {Phys. Lett.}\ }\textbf {\bibinfo {volume}
  {B788}},\ \bibinfo {pages} {453} (\bibinfo {year} {2019}{\natexlab{a}})},\
  \Eprint {http://arxiv.org/abs/1808.01437} {arXiv:1808.01437 [hep-ph]}
  \BibitemShut {NoStop}%
\bibitem [{\citenamefont {Radyushkin}(2019)}]{Radyushkin:2018nbf}%
  \BibitemOpen
  \bibfield  {author} {\bibinfo {author} {\bibfnamefont {A.~V.}\ \bibnamefont
  {Radyushkin}},\ }\href {\doibase 10.1016/j.physletb.2018.11.047} {\bibfield
  {journal} {\bibinfo  {journal} {Phys. Lett.}\ }\textbf {\bibinfo {volume}
  {B788}},\ \bibinfo {pages} {380} (\bibinfo {year} {2019})},\ \Eprint
  {http://arxiv.org/abs/1807.07509} {arXiv:1807.07509 [hep-ph]} \BibitemShut
  {NoStop}%
\bibitem [{\citenamefont {Zhang}\ \emph
  {et~al.}(2019{\natexlab{a}})\citenamefont {Zhang}, \citenamefont {Ji},
  \citenamefont {Schäfer}, \citenamefont {Wang},\ and\ \citenamefont
  {Zhao}}]{Zhang:2018diq}%
  \BibitemOpen
  \bibfield  {author} {\bibinfo {author} {\bibfnamefont {J.-H.}\ \bibnamefont
  {Zhang}}, \bibinfo {author} {\bibfnamefont {X.}~\bibnamefont {Ji}}, \bibinfo
  {author} {\bibfnamefont {A.}~\bibnamefont {Schäfer}}, \bibinfo {author}
  {\bibfnamefont {W.}~\bibnamefont {Wang}}, \ and\ \bibinfo {author}
  {\bibfnamefont {S.}~\bibnamefont {Zhao}},\ }\href {\doibase
  10.1103/PhysRevLett.122.142001} {\bibfield  {journal} {\bibinfo  {journal}
  {Phys. Rev. Lett.}\ }\textbf {\bibinfo {volume} {122}},\ \bibinfo {pages}
  {142001} (\bibinfo {year} {2019}{\natexlab{a}})},\ \Eprint
  {http://arxiv.org/abs/1808.10824} {arXiv:1808.10824 [hep-ph]} \BibitemShut
  {NoStop}%
\bibitem [{\citenamefont {Li}\ \emph {et~al.}(2019)\citenamefont {Li},
  \citenamefont {Ma},\ and\ \citenamefont {Qiu}}]{Li:2018tpe}%
  \BibitemOpen
  \bibfield  {author} {\bibinfo {author} {\bibfnamefont {Z.-Y.}\ \bibnamefont
  {Li}}, \bibinfo {author} {\bibfnamefont {Y.-Q.}\ \bibnamefont {Ma}}, \ and\
  \bibinfo {author} {\bibfnamefont {J.-W.}\ \bibnamefont {Qiu}},\ }\href
  {\doibase 10.1103/PhysRevLett.122.062002} {\bibfield  {journal} {\bibinfo
  {journal} {Phys. Rev. Lett.}\ }\textbf {\bibinfo {volume} {122}},\ \bibinfo
  {pages} {062002} (\bibinfo {year} {2019})},\ \Eprint
  {http://arxiv.org/abs/1809.01836} {arXiv:1809.01836 [hep-ph]} \BibitemShut
  {NoStop}%
\bibitem [{\citenamefont {Braun}\ \emph {et~al.}(2019)\citenamefont {Braun},
  \citenamefont {Vladimirov},\ and\ \citenamefont {Zhang}}]{Braun:2018brg}%
  \BibitemOpen
  \bibfield  {author} {\bibinfo {author} {\bibfnamefont {V.~M.}\ \bibnamefont
  {Braun}}, \bibinfo {author} {\bibfnamefont {A.}~\bibnamefont {Vladimirov}}, \
  and\ \bibinfo {author} {\bibfnamefont {J.-H.}\ \bibnamefont {Zhang}},\ }\href
  {\doibase 10.1103/PhysRevD.99.014013} {\bibfield  {journal} {\bibinfo
  {journal} {Phys. Rev.}\ }\textbf {\bibinfo {volume} {D99}},\ \bibinfo {pages}
  {014013} (\bibinfo {year} {2019})},\ \Eprint
  {http://arxiv.org/abs/1810.00048} {arXiv:1810.00048 [hep-ph]} \BibitemShut
  {NoStop}%
\bibitem [{\citenamefont {Lin}\ \emph {et~al.}(2015)\citenamefont {Lin},
  \citenamefont {Chen}, \citenamefont {Cohen},\ and\ \citenamefont
  {Ji}}]{Lin:2014zya}%
  \BibitemOpen
  \bibfield  {author} {\bibinfo {author} {\bibfnamefont {H.-W.}\ \bibnamefont
  {Lin}}, \bibinfo {author} {\bibfnamefont {J.-W.}\ \bibnamefont {Chen}},
  \bibinfo {author} {\bibfnamefont {S.~D.}\ \bibnamefont {Cohen}}, \ and\
  \bibinfo {author} {\bibfnamefont {X.}~\bibnamefont {Ji}},\ }\href {\doibase
  10.1103/PhysRevD.91.054510} {\bibfield  {journal} {\bibinfo  {journal} {Phys.
  Rev.}\ }\textbf {\bibinfo {volume} {D91}},\ \bibinfo {pages} {054510}
  (\bibinfo {year} {2015})},\ \Eprint {http://arxiv.org/abs/1402.1462}
  {arXiv:1402.1462 [hep-ph]} \BibitemShut {NoStop}%
\bibitem [{\citenamefont {Alexandrou}\ \emph {et~al.}(2015)\citenamefont
  {Alexandrou}, \citenamefont {Cichy}, \citenamefont {Drach}, \citenamefont
  {Garcia-Ramos}, \citenamefont {Hadjiyiannakou}, \citenamefont {Jansen},
  \citenamefont {Steffens},\ and\ \citenamefont {Wiese}}]{Alexandrou:2015rja}%
  \BibitemOpen
  \bibfield  {author} {\bibinfo {author} {\bibfnamefont {C.}~\bibnamefont
  {Alexandrou}}, \bibinfo {author} {\bibfnamefont {K.}~\bibnamefont {Cichy}},
  \bibinfo {author} {\bibfnamefont {V.}~\bibnamefont {Drach}}, \bibinfo
  {author} {\bibfnamefont {E.}~\bibnamefont {Garcia-Ramos}}, \bibinfo {author}
  {\bibfnamefont {K.}~\bibnamefont {Hadjiyiannakou}}, \bibinfo {author}
  {\bibfnamefont {K.}~\bibnamefont {Jansen}}, \bibinfo {author} {\bibfnamefont
  {F.}~\bibnamefont {Steffens}}, \ and\ \bibinfo {author} {\bibfnamefont
  {C.}~\bibnamefont {Wiese}},\ }\href {\doibase 10.1103/PhysRevD.92.014502}
  {\bibfield  {journal} {\bibinfo  {journal} {Phys. Rev.}\ }\textbf {\bibinfo
  {volume} {D92}},\ \bibinfo {pages} {014502} (\bibinfo {year} {2015})},\
  \Eprint {http://arxiv.org/abs/1504.07455} {arXiv:1504.07455 [hep-lat]}
  \BibitemShut {NoStop}%
\bibitem [{\citenamefont {Alexandrou}\ \emph
  {et~al.}(2017{\natexlab{b}})\citenamefont {Alexandrou}, \citenamefont
  {Cichy}, \citenamefont {Constantinou}, \citenamefont {Hadjiyiannakou},
  \citenamefont {Jansen}, \citenamefont {Steffens},\ and\ \citenamefont
  {Wiese}}]{Alexandrou:2016jqi}%
  \BibitemOpen
  \bibfield  {author} {\bibinfo {author} {\bibfnamefont {C.}~\bibnamefont
  {Alexandrou}}, \bibinfo {author} {\bibfnamefont {K.}~\bibnamefont {Cichy}},
  \bibinfo {author} {\bibfnamefont {M.}~\bibnamefont {Constantinou}}, \bibinfo
  {author} {\bibfnamefont {K.}~\bibnamefont {Hadjiyiannakou}}, \bibinfo
  {author} {\bibfnamefont {K.}~\bibnamefont {Jansen}}, \bibinfo {author}
  {\bibfnamefont {F.}~\bibnamefont {Steffens}}, \ and\ \bibinfo {author}
  {\bibfnamefont {C.}~\bibnamefont {Wiese}},\ }\href {\doibase
  10.1103/PhysRevD.96.014513} {\bibfield  {journal} {\bibinfo  {journal} {Phys.
  Rev.}\ }\textbf {\bibinfo {volume} {D96}},\ \bibinfo {pages} {014513}
  (\bibinfo {year} {2017}{\natexlab{b}})},\ \Eprint
  {http://arxiv.org/abs/1610.03689} {arXiv:1610.03689 [hep-lat]} \BibitemShut
  {NoStop}%
\bibitem [{\citenamefont {Zhang}\ \emph
  {et~al.}(2019{\natexlab{b}})\citenamefont {Zhang}, \citenamefont {Jin},
  \citenamefont {Lin}, \citenamefont {Schäfer}, \citenamefont {Sun},
  \citenamefont {Yang}, \citenamefont {Zhang}, \citenamefont {Zhao},\ and\
  \citenamefont {Chen}}]{Chen:2017gck}%
  \BibitemOpen
  \bibfield  {author} {\bibinfo {author} {\bibfnamefont {J.-H.}\ \bibnamefont
  {Zhang}}, \bibinfo {author} {\bibfnamefont {L.}~\bibnamefont {Jin}}, \bibinfo
  {author} {\bibfnamefont {H.-W.}\ \bibnamefont {Lin}}, \bibinfo {author}
  {\bibfnamefont {A.}~\bibnamefont {Schäfer}}, \bibinfo {author}
  {\bibfnamefont {P.}~\bibnamefont {Sun}}, \bibinfo {author} {\bibfnamefont
  {Y.-B.}\ \bibnamefont {Yang}}, \bibinfo {author} {\bibfnamefont
  {R.}~\bibnamefont {Zhang}}, \bibinfo {author} {\bibfnamefont
  {Y.}~\bibnamefont {Zhao}}, \ and\ \bibinfo {author} {\bibfnamefont {J.-W.}\
  \bibnamefont {Chen}} (\bibinfo {collaboration} {LP3}),\ }\href {\doibase
  10.1016/j.nuclphysb.2018.12.020} {\bibfield  {journal} {\bibinfo  {journal}
  {Nucl. Phys.}\ }\textbf {\bibinfo {volume} {B939}},\ \bibinfo {pages} {429}
  (\bibinfo {year} {2019}{\natexlab{b}})},\ \Eprint
  {http://arxiv.org/abs/1712.10025} {arXiv:1712.10025 [hep-ph]} \BibitemShut
  {NoStop}%
\bibitem [{\citenamefont {Alexandrou}\ \emph
  {et~al.}(2018{\natexlab{a}})\citenamefont {Alexandrou}, \citenamefont
  {Cichy}, \citenamefont {Constantinou}, \citenamefont {Jansen}, \citenamefont
  {Scapellato},\ and\ \citenamefont {Steffens}}]{Alexandrou:2018pbm}%
  \BibitemOpen
  \bibfield  {author} {\bibinfo {author} {\bibfnamefont {C.}~\bibnamefont
  {Alexandrou}}, \bibinfo {author} {\bibfnamefont {K.}~\bibnamefont {Cichy}},
  \bibinfo {author} {\bibfnamefont {M.}~\bibnamefont {Constantinou}}, \bibinfo
  {author} {\bibfnamefont {K.}~\bibnamefont {Jansen}}, \bibinfo {author}
  {\bibfnamefont {A.}~\bibnamefont {Scapellato}}, \ and\ \bibinfo {author}
  {\bibfnamefont {F.}~\bibnamefont {Steffens}},\ }\href {\doibase
  10.1103/PhysRevLett.121.112001} {\bibfield  {journal} {\bibinfo  {journal}
  {Phys. Rev. Lett.}\ }\textbf {\bibinfo {volume} {121}},\ \bibinfo {pages}
  {112001} (\bibinfo {year} {2018}{\natexlab{a}})},\ \Eprint
  {http://arxiv.org/abs/1803.02685} {arXiv:1803.02685 [hep-lat]} \BibitemShut
  {NoStop}%
\bibitem [{\citenamefont {Chen}\ \emph
  {et~al.}(2018{\natexlab{b}})\citenamefont {Chen}, \citenamefont {Jin},
  \citenamefont {Lin}, \citenamefont {Liu}, \citenamefont {Yang}, \citenamefont
  {Zhang},\ and\ \citenamefont {Zhao}}]{Chen:2018xof}%
  \BibitemOpen
  \bibfield  {author} {\bibinfo {author} {\bibfnamefont {J.-W.}\ \bibnamefont
  {Chen}}, \bibinfo {author} {\bibfnamefont {L.}~\bibnamefont {Jin}}, \bibinfo
  {author} {\bibfnamefont {H.-W.}\ \bibnamefont {Lin}}, \bibinfo {author}
  {\bibfnamefont {Y.-S.}\ \bibnamefont {Liu}}, \bibinfo {author} {\bibfnamefont
  {Y.-B.}\ \bibnamefont {Yang}}, \bibinfo {author} {\bibfnamefont {J.-H.}\
  \bibnamefont {Zhang}}, \ and\ \bibinfo {author} {\bibfnamefont
  {Y.}~\bibnamefont {Zhao}},\ }\href@noop {} {\  (\bibinfo {year}
  {2018}{\natexlab{b}})},\ \Eprint {http://arxiv.org/abs/1803.04393}
  {arXiv:1803.04393 [hep-lat]} \BibitemShut {NoStop}%
\bibitem [{\citenamefont {Chen}\ \emph
  {et~al.}(2018{\natexlab{c}})\citenamefont {Chen}, \citenamefont {Jin},
  \citenamefont {Lin}, \citenamefont {Liu}, \citenamefont {Schäfer},
  \citenamefont {Yang}, \citenamefont {Zhang},\ and\ \citenamefont
  {Zhao}}]{Chen:2018fwa}%
  \BibitemOpen
  \bibfield  {author} {\bibinfo {author} {\bibfnamefont {J.-W.}\ \bibnamefont
  {Chen}}, \bibinfo {author} {\bibfnamefont {L.}~\bibnamefont {Jin}}, \bibinfo
  {author} {\bibfnamefont {H.-W.}\ \bibnamefont {Lin}}, \bibinfo {author}
  {\bibfnamefont {Y.-S.}\ \bibnamefont {Liu}}, \bibinfo {author} {\bibfnamefont
  {A.}~\bibnamefont {Schäfer}}, \bibinfo {author} {\bibfnamefont {Y.-B.}\
  \bibnamefont {Yang}}, \bibinfo {author} {\bibfnamefont {J.-H.}\ \bibnamefont
  {Zhang}}, \ and\ \bibinfo {author} {\bibfnamefont {Y.}~\bibnamefont {Zhao}},\
  }\href@noop {} {\  (\bibinfo {year} {2018}{\natexlab{c}})},\ \Eprint
  {http://arxiv.org/abs/1804.01483} {arXiv:1804.01483 [hep-lat]} \BibitemShut
  {NoStop}%
\bibitem [{\citenamefont {Alexandrou}\ \emph
  {et~al.}(2018{\natexlab{b}})\citenamefont {Alexandrou}, \citenamefont
  {Cichy}, \citenamefont {Constantinou}, \citenamefont {Jansen}, \citenamefont
  {Scapellato},\ and\ \citenamefont {Steffens}}]{Alexandrou:2018eet}%
  \BibitemOpen
  \bibfield  {author} {\bibinfo {author} {\bibfnamefont {C.}~\bibnamefont
  {Alexandrou}}, \bibinfo {author} {\bibfnamefont {K.}~\bibnamefont {Cichy}},
  \bibinfo {author} {\bibfnamefont {M.}~\bibnamefont {Constantinou}}, \bibinfo
  {author} {\bibfnamefont {K.}~\bibnamefont {Jansen}}, \bibinfo {author}
  {\bibfnamefont {A.}~\bibnamefont {Scapellato}}, \ and\ \bibinfo {author}
  {\bibfnamefont {F.}~\bibnamefont {Steffens}},\ }\href {\doibase
  10.1103/PhysRevD.98.091503} {\bibfield  {journal} {\bibinfo  {journal} {Phys.
  Rev.}\ }\textbf {\bibinfo {volume} {D98}},\ \bibinfo {pages} {091503}
  (\bibinfo {year} {2018}{\natexlab{b}})},\ \Eprint
  {http://arxiv.org/abs/1807.00232} {arXiv:1807.00232 [hep-lat]} \BibitemShut
  {NoStop}%
\bibitem [{\citenamefont {Lin}\ \emph {et~al.}(2018{\natexlab{b}})\citenamefont
  {Lin}, \citenamefont {Chen}, \citenamefont {Ji}, \citenamefont {Jin},
  \citenamefont {Li}, \citenamefont {Liu}, \citenamefont {Yang}, \citenamefont
  {Zhang},\ and\ \citenamefont {Zhao}}]{Lin:2018qky}%
  \BibitemOpen
  \bibfield  {author} {\bibinfo {author} {\bibfnamefont {H.-W.}\ \bibnamefont
  {Lin}}, \bibinfo {author} {\bibfnamefont {J.-W.}\ \bibnamefont {Chen}},
  \bibinfo {author} {\bibfnamefont {X.}~\bibnamefont {Ji}}, \bibinfo {author}
  {\bibfnamefont {L.}~\bibnamefont {Jin}}, \bibinfo {author} {\bibfnamefont
  {R.}~\bibnamefont {Li}}, \bibinfo {author} {\bibfnamefont {Y.-S.}\
  \bibnamefont {Liu}}, \bibinfo {author} {\bibfnamefont {Y.-B.}\ \bibnamefont
  {Yang}}, \bibinfo {author} {\bibfnamefont {J.-H.}\ \bibnamefont {Zhang}}, \
  and\ \bibinfo {author} {\bibfnamefont {Y.}~\bibnamefont {Zhao}},\ }\href
  {\doibase 10.1103/PhysRevLett.121.242003} {\bibfield  {journal} {\bibinfo
  {journal} {Phys. Rev. Lett.}\ }\textbf {\bibinfo {volume} {121}},\ \bibinfo
  {pages} {242003} (\bibinfo {year} {2018}{\natexlab{b}})},\ \Eprint
  {http://arxiv.org/abs/1807.07431} {arXiv:1807.07431 [hep-lat]} \BibitemShut
  {NoStop}%
\bibitem [{\citenamefont {Fan}\ \emph {et~al.}(2018)\citenamefont {Fan},
  \citenamefont {Yang}, \citenamefont {Anthony}, \citenamefont {Lin},\ and\
  \citenamefont {Liu}}]{Fan:2018dxu}%
  \BibitemOpen
  \bibfield  {author} {\bibinfo {author} {\bibfnamefont {Z.-Y.}\ \bibnamefont
  {Fan}}, \bibinfo {author} {\bibfnamefont {Y.-B.}\ \bibnamefont {Yang}},
  \bibinfo {author} {\bibfnamefont {A.}~\bibnamefont {Anthony}}, \bibinfo
  {author} {\bibfnamefont {H.-W.}\ \bibnamefont {Lin}}, \ and\ \bibinfo
  {author} {\bibfnamefont {K.-F.}\ \bibnamefont {Liu}},\ }\href {\doibase
  10.1103/PhysRevLett.121.242001} {\bibfield  {journal} {\bibinfo  {journal}
  {Phys. Rev. Lett.}\ }\textbf {\bibinfo {volume} {121}},\ \bibinfo {pages}
  {242001} (\bibinfo {year} {2018})},\ \Eprint
  {http://arxiv.org/abs/1808.02077} {arXiv:1808.02077 [hep-lat]} \BibitemShut
  {NoStop}%
\bibitem [{\citenamefont {Liu}\ \emph {et~al.}(2018{\natexlab{b}})\citenamefont
  {Liu}, \citenamefont {Chen}, \citenamefont {Jin}, \citenamefont {Li},
  \citenamefont {Lin}, \citenamefont {Yang}, \citenamefont {Zhang},\ and\
  \citenamefont {Zhao}}]{Liu:2018hxv}%
  \BibitemOpen
  \bibfield  {author} {\bibinfo {author} {\bibfnamefont {Y.-S.}\ \bibnamefont
  {Liu}}, \bibinfo {author} {\bibfnamefont {J.-W.}\ \bibnamefont {Chen}},
  \bibinfo {author} {\bibfnamefont {L.}~\bibnamefont {Jin}}, \bibinfo {author}
  {\bibfnamefont {R.}~\bibnamefont {Li}}, \bibinfo {author} {\bibfnamefont
  {H.-W.}\ \bibnamefont {Lin}}, \bibinfo {author} {\bibfnamefont {Y.-B.}\
  \bibnamefont {Yang}}, \bibinfo {author} {\bibfnamefont {J.-H.}\ \bibnamefont
  {Zhang}}, \ and\ \bibinfo {author} {\bibfnamefont {Y.}~\bibnamefont {Zhao}},\
  }\href@noop {} {\  (\bibinfo {year} {2018}{\natexlab{b}})},\ \Eprint
  {http://arxiv.org/abs/1810.05043} {arXiv:1810.05043 [hep-lat]} \BibitemShut
  {NoStop}%
\bibitem [{\citenamefont {Wang}\ \emph {et~al.}(2019)\citenamefont {Wang},
  \citenamefont {Zhang}, \citenamefont {Zhao},\ and\ \citenamefont
  {Zhu}}]{Wang:2019tgg}%
  \BibitemOpen
  \bibfield  {author} {\bibinfo {author} {\bibfnamefont {W.}~\bibnamefont
  {Wang}}, \bibinfo {author} {\bibfnamefont {J.-H.}\ \bibnamefont {Zhang}},
  \bibinfo {author} {\bibfnamefont {S.}~\bibnamefont {Zhao}}, \ and\ \bibinfo
  {author} {\bibfnamefont {R.}~\bibnamefont {Zhu}},\ }\href@noop {} {\
  (\bibinfo {year} {2019})},\ \Eprint {http://arxiv.org/abs/1904.00978}
  {arXiv:1904.00978 [hep-ph]} \BibitemShut {NoStop}%
\bibitem [{\citenamefont {Dulat}\ \emph {et~al.}(2016)\citenamefont {Dulat},
  \citenamefont {Hou}, \citenamefont {Gao}, \citenamefont {Guzzi},
  \citenamefont {Huston}, \citenamefont {Nadolsky}, \citenamefont {Pumplin},
  \citenamefont {Schmidt}, \citenamefont {Stump},\ and\ \citenamefont
  {Yuan}}]{Dulat:2015mca}%
  \BibitemOpen
  \bibfield  {author} {\bibinfo {author} {\bibfnamefont {S.}~\bibnamefont
  {Dulat}}, \bibinfo {author} {\bibfnamefont {T.-J.}\ \bibnamefont {Hou}},
  \bibinfo {author} {\bibfnamefont {J.}~\bibnamefont {Gao}}, \bibinfo {author}
  {\bibfnamefont {M.}~\bibnamefont {Guzzi}}, \bibinfo {author} {\bibfnamefont
  {J.}~\bibnamefont {Huston}}, \bibinfo {author} {\bibfnamefont
  {P.}~\bibnamefont {Nadolsky}}, \bibinfo {author} {\bibfnamefont
  {J.}~\bibnamefont {Pumplin}}, \bibinfo {author} {\bibfnamefont
  {C.}~\bibnamefont {Schmidt}}, \bibinfo {author} {\bibfnamefont
  {D.}~\bibnamefont {Stump}}, \ and\ \bibinfo {author} {\bibfnamefont {C.~P.}\
  \bibnamefont {Yuan}},\ }\href {\doibase 10.1103/PhysRevD.93.033006}
  {\bibfield  {journal} {\bibinfo  {journal} {Phys. Rev.}\ }\textbf {\bibinfo
  {volume} {D93}},\ \bibinfo {pages} {033006} (\bibinfo {year} {2016})},\
  \Eprint {http://arxiv.org/abs/1506.07443} {arXiv:1506.07443 [hep-ph]}
  \BibitemShut {NoStop}%
\bibitem [{\citenamefont {Ball}\ \emph {et~al.}(2017)\citenamefont {Ball} \emph
  {et~al.}}]{Ball:2017nwa}%
  \BibitemOpen
  \bibfield  {author} {\bibinfo {author} {\bibfnamefont {R.~D.}\ \bibnamefont
  {Ball}} \emph {et~al.} (\bibinfo {collaboration} {NNPDF}),\ }\href {\doibase
  10.1140/epjc/s10052-017-5199-5} {\bibfield  {journal} {\bibinfo  {journal}
  {Eur. Phys. J.}\ }\textbf {\bibinfo {volume} {C77}},\ \bibinfo {pages} {663}
  (\bibinfo {year} {2017})},\ \Eprint {http://arxiv.org/abs/1706.00428}
  {arXiv:1706.00428 [hep-ph]} \BibitemShut {NoStop}%
\bibitem [{\citenamefont {Harland-Lang}\ \emph {et~al.}(2015)\citenamefont
  {Harland-Lang}, \citenamefont {Martin}, \citenamefont {Motylinski},\ and\
  \citenamefont {Thorne}}]{Harland-Lang:2014zoa}%
  \BibitemOpen
  \bibfield  {author} {\bibinfo {author} {\bibfnamefont {L.~A.}\ \bibnamefont
  {Harland-Lang}}, \bibinfo {author} {\bibfnamefont {A.~D.}\ \bibnamefont
  {Martin}}, \bibinfo {author} {\bibfnamefont {P.}~\bibnamefont {Motylinski}},
  \ and\ \bibinfo {author} {\bibfnamefont {R.~S.}\ \bibnamefont {Thorne}},\
  }\href {\doibase 10.1140/epjc/s10052-015-3397-6} {\bibfield  {journal}
  {\bibinfo  {journal} {Eur. Phys. J.}\ }\textbf {\bibinfo {volume} {C75}},\
  \bibinfo {pages} {204} (\bibinfo {year} {2015})},\ \Eprint
  {http://arxiv.org/abs/1412.3989} {arXiv:1412.3989 [hep-ph]} \BibitemShut
  {NoStop}%
\bibitem [{\citenamefont {Nocera}\ \emph {et~al.}(2014)\citenamefont {Nocera},
  \citenamefont {Ball}, \citenamefont {Forte}, \citenamefont {Ridolfi},\ and\
  \citenamefont {Rojo}}]{Nocera:2014gqa}%
  \BibitemOpen
  \bibfield  {author} {\bibinfo {author} {\bibfnamefont {E.~R.}\ \bibnamefont
  {Nocera}}, \bibinfo {author} {\bibfnamefont {R.~D.}\ \bibnamefont {Ball}},
  \bibinfo {author} {\bibfnamefont {S.}~\bibnamefont {Forte}}, \bibinfo
  {author} {\bibfnamefont {G.}~\bibnamefont {Ridolfi}}, \ and\ \bibinfo
  {author} {\bibfnamefont {J.}~\bibnamefont {Rojo}} (\bibinfo {collaboration}
  {NNPDF}),\ }\href {\doibase 10.1016/j.nuclphysb.2014.08.008} {\bibfield
  {journal} {\bibinfo  {journal} {Nucl. Phys.}\ }\textbf {\bibinfo {volume}
  {B887}},\ \bibinfo {pages} {276} (\bibinfo {year} {2014})},\ \Eprint
  {http://arxiv.org/abs/1406.5539} {arXiv:1406.5539 [hep-ph]} \BibitemShut
  {NoStop}%
\bibitem [{\citenamefont {Ethier}\ \emph {et~al.}(2017)\citenamefont {Ethier},
  \citenamefont {Sato},\ and\ \citenamefont {Melnitchouk}}]{Ethier:2017zbq}%
  \BibitemOpen
  \bibfield  {author} {\bibinfo {author} {\bibfnamefont {J.~J.}\ \bibnamefont
  {Ethier}}, \bibinfo {author} {\bibfnamefont {N.}~\bibnamefont {Sato}}, \ and\
  \bibinfo {author} {\bibfnamefont {W.}~\bibnamefont {Melnitchouk}},\ }\href
  {\doibase 10.1103/PhysRevLett.119.132001} {\bibfield  {journal} {\bibinfo
  {journal} {Phys. Rev. Lett.}\ }\textbf {\bibinfo {volume} {119}},\ \bibinfo
  {pages} {132001} (\bibinfo {year} {2017})},\ \Eprint
  {http://arxiv.org/abs/1705.05889} {arXiv:1705.05889 [hep-ph]} \BibitemShut
  {NoStop}%
\bibitem [{\citenamefont {Bhattacharya}\ \emph
  {et~al.}(2019{\natexlab{b}})\citenamefont {Bhattacharya}, \citenamefont
  {Cocuzza},\ and\ \citenamefont {Metz}}]{Bhattacharya:2019cme}%
  \BibitemOpen
  \bibfield  {author} {\bibinfo {author} {\bibfnamefont {S.}~\bibnamefont
  {Bhattacharya}}, \bibinfo {author} {\bibfnamefont {C.}~\bibnamefont
  {Cocuzza}}, \ and\ \bibinfo {author} {\bibfnamefont {A.}~\bibnamefont
  {Metz}},\ }\href@noop {} {\  (\bibinfo {year} {2019}{\natexlab{b}})},\
  \Eprint {http://arxiv.org/abs/1903.05721} {arXiv:1903.05721 [hep-ph]}
  \BibitemShut {NoStop}%
\bibitem [{\citenamefont {Burkardt}(2003)}]{Burkardt:2002hr}%
  \BibitemOpen
  \bibfield  {author} {\bibinfo {author} {\bibfnamefont {M.}~\bibnamefont
  {Burkardt}},\ }\href {\doibase 10.1142/S0217751X03012370} {\bibfield
  {journal} {\bibinfo  {journal} {Int. J. Mod. Phys.}\ }\textbf {\bibinfo
  {volume} {A18}},\ \bibinfo {pages} {173} (\bibinfo {year} {2003})},\ \Eprint
  {http://arxiv.org/abs/hep-ph/0207047} {arXiv:hep-ph/0207047 [hep-ph]}
  \BibitemShut {NoStop}%
\bibitem [{\citenamefont {Ralston}\ and\ \citenamefont
  {Pire}(2002)}]{Ralston:2001xs}%
  \BibitemOpen
  \bibfield  {author} {\bibinfo {author} {\bibfnamefont {J.~P.}\ \bibnamefont
  {Ralston}}\ and\ \bibinfo {author} {\bibfnamefont {B.}~\bibnamefont {Pire}},\
  }\href {\doibase 10.1103/PhysRevD.66.111501} {\bibfield  {journal} {\bibinfo
  {journal} {Phys. Rev.}\ }\textbf {\bibinfo {volume} {D66}},\ \bibinfo {pages}
  {111501} (\bibinfo {year} {2002})},\ \Eprint
  {http://arxiv.org/abs/hep-ph/0110075} {arXiv:hep-ph/0110075 [hep-ph]}
  \BibitemShut {NoStop}%
\bibitem [{\citenamefont {Follana}\ \emph {et~al.}(2007)\citenamefont
  {Follana}, \citenamefont {Mason}, \citenamefont {Davies}, \citenamefont
  {Hornbostel}, \citenamefont {Lepage}, \citenamefont {Shigemitsu},
  \citenamefont {Trottier},\ and\ \citenamefont {Wong}}]{Follana:2006rc}%
  \BibitemOpen
  \bibfield  {author} {\bibinfo {author} {\bibfnamefont {E.}~\bibnamefont
  {Follana}}, \bibinfo {author} {\bibfnamefont {Q.}~\bibnamefont {Mason}},
  \bibinfo {author} {\bibfnamefont {C.}~\bibnamefont {Davies}}, \bibinfo
  {author} {\bibfnamefont {K.}~\bibnamefont {Hornbostel}}, \bibinfo {author}
  {\bibfnamefont {G.~P.}\ \bibnamefont {Lepage}}, \bibinfo {author}
  {\bibfnamefont {J.}~\bibnamefont {Shigemitsu}}, \bibinfo {author}
  {\bibfnamefont {H.}~\bibnamefont {Trottier}}, \ and\ \bibinfo {author}
  {\bibfnamefont {K.}~\bibnamefont {Wong}} (\bibinfo {collaboration} {HPQCD,
  UKQCD}),\ }\href {\doibase 10.1103/PhysRevD.75.054502} {\bibfield  {journal}
  {\bibinfo  {journal} {Phys. Rev.}\ }\textbf {\bibinfo {volume} {D75}},\
  \bibinfo {pages} {054502} (\bibinfo {year} {2007})},\ \Eprint
  {http://arxiv.org/abs/hep-lat/0610092} {arXiv:hep-lat/0610092 [hep-lat]}
  \BibitemShut {NoStop}%
\bibitem [{\citenamefont {Bazavov}\ \emph {et~al.}(2013)\citenamefont {Bazavov}
  \emph {et~al.}}]{Bazavov:2012xda}%
  \BibitemOpen
  \bibfield  {author} {\bibinfo {author} {\bibfnamefont {A.}~\bibnamefont
  {Bazavov}} \emph {et~al.} (\bibinfo {collaboration} {MILC}),\ }\href
  {\doibase 10.1103/PhysRevD.87.054505} {\bibfield  {journal} {\bibinfo
  {journal} {Phys. Rev.}\ }\textbf {\bibinfo {volume} {D87}},\ \bibinfo {pages}
  {054505} (\bibinfo {year} {2013})},\ \Eprint {http://arxiv.org/abs/1212.4768}
  {arXiv:1212.4768 [hep-lat]} \BibitemShut {NoStop}%
\bibitem [{\citenamefont {Burkardt}(2000)}]{Burkardt:2000za}%
  \BibitemOpen
  \bibfield  {author} {\bibinfo {author} {\bibfnamefont {M.}~\bibnamefont
  {Burkardt}},\ }\href {\doibase 10.1103/PhysRevD.62.071503,
  10.1103/PhysRevD.66.119903} {\bibfield  {journal} {\bibinfo  {journal} {Phys.
  Rev.}\ }\textbf {\bibinfo {volume} {D62}},\ \bibinfo {pages} {071503}
  (\bibinfo {year} {2000})},\ \bibinfo {note} {[Erratum: Phys.
  Rev.D66,119903(2002)]},\ \Eprint {http://arxiv.org/abs/hep-ph/0005108}
  {arXiv:hep-ph/0005108 [hep-ph]} \BibitemShut {NoStop}%
\bibitem [{\citenamefont {Hasenfratz}\ and\ \citenamefont
  {Knechtli}(2001)}]{Hasenfratz:2001hp}%
  \BibitemOpen
  \bibfield  {author} {\bibinfo {author} {\bibfnamefont {A.}~\bibnamefont
  {Hasenfratz}}\ and\ \bibinfo {author} {\bibfnamefont {F.}~\bibnamefont
  {Knechtli}},\ }\href {\doibase 10.1103/PhysRevD.64.034504} {\bibfield
  {journal} {\bibinfo  {journal} {Phys. Rev.}\ }\textbf {\bibinfo {volume}
  {D64}},\ \bibinfo {pages} {034504} (\bibinfo {year} {2001})},\ \Eprint
  {http://arxiv.org/abs/hep-lat/0103029} {arXiv:hep-lat/0103029 [hep-lat]}
  \BibitemShut {NoStop}%
\bibitem [{\citenamefont {Gupta}\ \emph {et~al.}(2017)\citenamefont {Gupta},
  \citenamefont {Jang}, \citenamefont {Lin}, \citenamefont {Yoon},\ and\
  \citenamefont {Bhattacharya}}]{Rajan:2017lxk}%
  \BibitemOpen
  \bibfield  {author} {\bibinfo {author} {\bibfnamefont {R.}~\bibnamefont
  {Gupta}}, \bibinfo {author} {\bibfnamefont {Y.-C.}\ \bibnamefont {Jang}},
  \bibinfo {author} {\bibfnamefont {H.-W.}\ \bibnamefont {Lin}}, \bibinfo
  {author} {\bibfnamefont {B.}~\bibnamefont {Yoon}}, \ and\ \bibinfo {author}
  {\bibfnamefont {T.}~\bibnamefont {Bhattacharya}},\ }\href {\doibase
  10.1103/PhysRevD.96.114503} {\bibfield  {journal} {\bibinfo  {journal} {Phys.
  Rev.}\ }\textbf {\bibinfo {volume} {D96}},\ \bibinfo {pages} {114503}
  (\bibinfo {year} {2017})},\ \Eprint {http://arxiv.org/abs/1705.06834}
  {arXiv:1705.06834 [hep-lat]} \BibitemShut {NoStop}%
\bibitem [{\citenamefont {Bhattacharya}\ \emph
  {et~al.}(2015{\natexlab{a}})\citenamefont {Bhattacharya}, \citenamefont
  {Cirigliano}, \citenamefont {Cohen}, \citenamefont {Gupta}, \citenamefont
  {Joseph}, \citenamefont {Lin},\ and\ \citenamefont
  {Yoon}}]{Bhattacharya:2015wna}%
  \BibitemOpen
  \bibfield  {author} {\bibinfo {author} {\bibfnamefont {T.}~\bibnamefont
  {Bhattacharya}}, \bibinfo {author} {\bibfnamefont {V.}~\bibnamefont
  {Cirigliano}}, \bibinfo {author} {\bibfnamefont {S.}~\bibnamefont {Cohen}},
  \bibinfo {author} {\bibfnamefont {R.}~\bibnamefont {Gupta}}, \bibinfo
  {author} {\bibfnamefont {A.}~\bibnamefont {Joseph}}, \bibinfo {author}
  {\bibfnamefont {H.-W.}\ \bibnamefont {Lin}}, \ and\ \bibinfo {author}
  {\bibfnamefont {B.}~\bibnamefont {Yoon}} (\bibinfo {collaboration} {PNDME}),\
  }\href {\doibase 10.1103/PhysRevD.92.094511} {\bibfield  {journal} {\bibinfo
  {journal} {Phys. Rev.}\ }\textbf {\bibinfo {volume} {D92}},\ \bibinfo {pages}
  {094511} (\bibinfo {year} {2015}{\natexlab{a}})},\ \Eprint
  {http://arxiv.org/abs/1506.06411} {arXiv:1506.06411 [hep-lat]} \BibitemShut
  {NoStop}%
\bibitem [{\citenamefont {Bhattacharya}\ \emph
  {et~al.}(2015{\natexlab{b}})\citenamefont {Bhattacharya}, \citenamefont
  {Cirigliano}, \citenamefont {Gupta}, \citenamefont {Lin},\ and\ \citenamefont
  {Yoon}}]{Bhattacharya:2015esa}%
  \BibitemOpen
  \bibfield  {author} {\bibinfo {author} {\bibfnamefont {T.}~\bibnamefont
  {Bhattacharya}}, \bibinfo {author} {\bibfnamefont {V.}~\bibnamefont
  {Cirigliano}}, \bibinfo {author} {\bibfnamefont {R.}~\bibnamefont {Gupta}},
  \bibinfo {author} {\bibfnamefont {H.-W.}\ \bibnamefont {Lin}}, \ and\
  \bibinfo {author} {\bibfnamefont {B.}~\bibnamefont {Yoon}},\ }\href {\doibase
  10.1103/PhysRevLett.115.212002} {\bibfield  {journal} {\bibinfo  {journal}
  {Phys. Rev. Lett.}\ }\textbf {\bibinfo {volume} {115}},\ \bibinfo {pages}
  {212002} (\bibinfo {year} {2015}{\natexlab{b}})},\ \Eprint
  {http://arxiv.org/abs/1506.04196} {arXiv:1506.04196 [hep-lat]} \BibitemShut
  {NoStop}%
\bibitem [{\citenamefont {Bhattacharya}\ \emph {et~al.}(2014)\citenamefont
  {Bhattacharya}, \citenamefont {Cohen}, \citenamefont {Gupta}, \citenamefont
  {Joseph}, \citenamefont {Lin},\ and\ \citenamefont
  {Yoon}}]{Bhattacharya:2013ehc}%
  \BibitemOpen
  \bibfield  {author} {\bibinfo {author} {\bibfnamefont {T.}~\bibnamefont
  {Bhattacharya}}, \bibinfo {author} {\bibfnamefont {S.~D.}\ \bibnamefont
  {Cohen}}, \bibinfo {author} {\bibfnamefont {R.}~\bibnamefont {Gupta}},
  \bibinfo {author} {\bibfnamefont {A.}~\bibnamefont {Joseph}}, \bibinfo
  {author} {\bibfnamefont {H.-W.}\ \bibnamefont {Lin}}, \ and\ \bibinfo
  {author} {\bibfnamefont {B.}~\bibnamefont {Yoon}},\ }\href {\doibase
  10.1103/PhysRevD.89.094502} {\bibfield  {journal} {\bibinfo  {journal} {Phys.
  Rev.}\ }\textbf {\bibinfo {volume} {D89}},\ \bibinfo {pages} {094502}
  (\bibinfo {year} {2014})},\ \Eprint {http://arxiv.org/abs/1306.5435}
  {arXiv:1306.5435 [hep-lat]} \BibitemShut {NoStop}%
\bibitem [{\citenamefont {Capitani}\ \emph {et~al.}(2012)\citenamefont
  {Capitani}, \citenamefont {Della~Morte}, \citenamefont {von Hippel},
  \citenamefont {Jager}, \citenamefont {Juttner}, \citenamefont {Knippschild},
  \citenamefont {Meyer},\ and\ \citenamefont {Wittig}}]{Capitani:2012gj}%
  \BibitemOpen
  \bibfield  {author} {\bibinfo {author} {\bibfnamefont {S.}~\bibnamefont
  {Capitani}}, \bibinfo {author} {\bibfnamefont {M.}~\bibnamefont
  {Della~Morte}}, \bibinfo {author} {\bibfnamefont {G.}~\bibnamefont {von
  Hippel}}, \bibinfo {author} {\bibfnamefont {B.}~\bibnamefont {Jager}},
  \bibinfo {author} {\bibfnamefont {A.}~\bibnamefont {Juttner}}, \bibinfo
  {author} {\bibfnamefont {B.}~\bibnamefont {Knippschild}}, \bibinfo {author}
  {\bibfnamefont {H.~B.}\ \bibnamefont {Meyer}}, \ and\ \bibinfo {author}
  {\bibfnamefont {H.}~\bibnamefont {Wittig}},\ }\href {\doibase
  10.1103/PhysRevD.86.074502} {\bibfield  {journal} {\bibinfo  {journal} {Phys.
  Rev.}\ }\textbf {\bibinfo {volume} {D86}},\ \bibinfo {pages} {074502}
  (\bibinfo {year} {2012})},\ \Eprint {http://arxiv.org/abs/1205.0180}
  {arXiv:1205.0180 [hep-lat]} \BibitemShut {NoStop}%
\bibitem [{\citenamefont {Brömmel}\ \emph {et~al.}(2007)\citenamefont
  {Brömmel} \emph {et~al.}}]{Brommel:2006ww}%
  \BibitemOpen
  \bibfield  {author} {\bibinfo {author} {\bibfnamefont {D.}~\bibnamefont
  {Brömmel}} \emph {et~al.} (\bibinfo {collaboration} {QCDSF/UKQCD}),\ }\href
  {\doibase 10.1140/epjc/s10052-007-0295-6} {\bibfield  {journal} {\bibinfo
  {journal} {Eur. Phys. J.}\ }\textbf {\bibinfo {volume} {C51}},\ \bibinfo
  {pages} {335} (\bibinfo {year} {2007})},\ \Eprint
  {http://arxiv.org/abs/hep-lat/0608021} {arXiv:hep-lat/0608021 [hep-lat]}
  \BibitemShut {NoStop}%
\bibitem [{\citenamefont {Diehl}\ \emph {et~al.}(2005)\citenamefont {Diehl},
  \citenamefont {Feldmann}, \citenamefont {Jakob},\ and\ \citenamefont
  {Kroll}}]{Diehl:2004cx}%
  \BibitemOpen
  \bibfield  {author} {\bibinfo {author} {\bibfnamefont {M.}~\bibnamefont
  {Diehl}}, \bibinfo {author} {\bibfnamefont {T.}~\bibnamefont {Feldmann}},
  \bibinfo {author} {\bibfnamefont {R.}~\bibnamefont {Jakob}}, \ and\ \bibinfo
  {author} {\bibfnamefont {P.}~\bibnamefont {Kroll}},\ }\href {\doibase
  10.1140/epjc/s2004-02063-4} {\bibfield  {journal} {\bibinfo  {journal} {Eur.
  Phys. J.}\ }\textbf {\bibinfo {volume} {C39}},\ \bibinfo {pages} {1}
  (\bibinfo {year} {2005})},\ \Eprint {http://arxiv.org/abs/hep-ph/0408173}
  {arXiv:hep-ph/0408173 [hep-ph]} \BibitemShut {NoStop}%
\bibitem [{\citenamefont {Edwards}\ and\ \citenamefont
  {Joo}(2005)}]{Edwards:2004sx}%
  \BibitemOpen
  \bibfield  {author} {\bibinfo {author} {\bibfnamefont {R.~G.}\ \bibnamefont
  {Edwards}}\ and\ \bibinfo {author} {\bibfnamefont {B.}~\bibnamefont {Joo}}
  (\bibinfo {collaboration} {SciDAC, LHPC, UKQCD}),\ }\bibfield  {booktitle}
  {\emph {\bibinfo {booktitle} {{Lattice field theory. Proceedings, 22nd
  International Symposium, Lattice 2004, Batavia, USA, June 21-26, 2004}}},\
  }\href {\doibase 10.1016/j.nuclphysbps.2004.11.254} {\bibfield  {journal}
  {\bibinfo  {journal} {Nucl. Phys. Proc. Suppl.}\ }\textbf {\bibinfo {volume}
  {140}},\ \bibinfo {pages} {832} (\bibinfo {year} {2005})},\ \bibinfo {note}
  {[,832(2004)]},\ \Eprint {http://arxiv.org/abs/hep-lat/0409003}
  {arXiv:hep-lat/0409003 [hep-lat]} \BibitemShut {NoStop}%
\end{thebibliography}%
\end{document}